\documentclass[twocolumn]{article}
\usepackage{arxiv}

\usepackage[utf8]{inputenc} 
\usepackage[T1]{fontenc}
\usepackage{amsmath, amssymb}
\usepackage{graphicx}
\usepackage{url}
\usepackage{hyperref}
\usepackage{booktabs}
\usepackage{nicefrac}
\usepackage{microtype}
\usepackage{lipsum} 
\usepackage{multicol}
\usepackage{multirow}

\title{Deep Image Prior Assisted ISAR Imaging For Missing Data Case}

\author{
  \normalsize Necmettin Bayar \\
  \normalsize Telecommunication Engineering Department\\
  \normalsize Istanbul Technical University\\
  \normalsize \texttt{bayarn20@itu.edu.tr}
  \and
  \normalsize Isin Erer \\
  \normalsize Telecommunication Engineering Department\\
  \normalsize Istanbul Technical University\\
  \normalsize \texttt{ierer@itu.edu.tr}
  \and
  \normalsize Deniz Kumlu \\
  \normalsize Telecommunication Engineering Department\\
  \normalsize Istanbul Technical University\\
  \normalsize \texttt{kumlud@itu.edu.tr}
}

\begin{document}

\twocolumn[
\begin{@twocolumnfalse}
\maketitle

\vspace{-1em}

\begin{abstract}
In Inverse Synthetic Aperture Radar (ISAR), random missing entries of the received radar echo matrix deteriorate the imaging quality, compromising target distinction from the background. Compressive sensing techniques or matrix completion prior to conventional imaging have been used in recent years to solve this issue. However, while the former techniques fail to preserve target continuity due to the sparsity constraint, the latter fails for high missing ratios. This paper proposes to use deep image prior (DIP) to complete the complex radar data and then obtain the radar image by conventional Fourier imaging. Real and imaginary parts are separately completed by independent deep structures and then put together for the imaging part. The proposed DIP based imaging method has been compared with IALM, 2D-SL0 and NNM methods visually and quantitatively for both simulated and real data. The results demonstrate an increase of 100\% for some extreme cases in terms of RMSE, 50\% increase on Correlation and 30\% increase on IC metrics quantitatively.
\end{abstract}

\vspace{1em}

\rule{\linewidth}{0.4pt} 
  \end{@twocolumnfalse}
]

\section{Introduction}
Inverse synthetic aperture radar (ISAR) imaging has crucial importance in many military applications such as detection, identification, and classification of targets.

However, it is widely known that traditional ISAR imaging based on the Fourier transform of the received backscattered data suffers from poor resolution, due to the fact that for most of the real world scenarios the data is collected from narrow angle and bandwidth.  Thus, in the last three decades, there have been many attempts to increase the resolution beyond the one provided by the Fourier transform. Spectral estimation based methods such as CAPON \cite{capon}, autoregressive modeling \cite{Ierer_AR,Ierer_AR2,Ierer_AR3}, MUSIC \cite{Music} etc. have widely used 2D radar images by using 2D Cartesian frequency data collected from a limited  observation angle and bandwidth area. However, these methods, besides needing priory information about the target such as the number of scattering centers, also require 2D modelling of the measurement data, thus have high complexity.

 Moreover, the low resolution problem caused by limited data, some problems that may occur during the measurements can seriously affect the quality of the resulting ISAR images. One of the fundamental problems in ISAR literature is the missing data case. Such a problem is not unique to ISAR, but is a problem that may be encountered in many measurements \cite{GNSS, Health, Fault}, and its elimination is essential to increase performance. This can arise from undesirable interference, external jamming signal, occlusion or some instrumental or receiving problems when obtaining the related data from a target. As a result, received data can be partially missing. Besides, the contemporary ISAR imaging systems use active phased array radar techniques and switch the radar beams in several directions to obtain multiple targets, which can  also lead to the missing data problem. Therefore, applying conventional Fourier transform to the obtained missing data usually results in corrupted or poor ISAR imaging results.

Sparse signal recovery algorithms which are based on compressive sensing (CS) theory \cite{CS_2006} can be used to solve this problem. According to CS theory, a sparse signal can be reconstructed  from a very few observed samples compared to that required by Nyquist sampling theorem. The important condition is to represent the signal sparsely by its redundant basis functions. In ISAR imaging, the spatial domain of a target is sparse thus CS techniques can be utilized for reconstruction purposes in the missing data case. Several methods which use sparse measurements in frequency, azimuth or in both domains have been introduced. In these approaches, the resulting optimization problem has been converted into 1D sparse reconstruction by stacking 2D measurement data into vector form. Due to the size of the data vector especially for large size ISAR images, a huge dictionary is required. A solution to decrease the computational load is 2D Smoothed L0 (2D-SL0) \cite{ghaffari_2009} which uses 2D data without any vectorization procedure. Although it has promising results, the method requires a separable sampling pattern, thus it fails for 2D random sampling case.

CS methods also assume that the
scatters are sparse. However, this assumption may not hold for complex radar targets, hence deterioration of the continuity of the target structures occurs in the recovered image. For the continuity of the object structures in the image, group or block sparsity approaches are proposed \cite{group_sparsity}, \cite{block_sparsity}. Ideally, groups should be defined according to the pixel group where the targets lie. However, since there is no a priori information on the scene, the groups are defined as neighbouring N × N pixels. This definition degrades the image quality, especially at the edges of the objects. Moreover, optimization step becomes more complicated, thus the group sparsity approach increases the computational load. 
An interesting alternative is \cite{Sonia_2016} where the sparse coefficients are not used to represent the imaging scene, but are multiplied by the dictionary to reconstruct the backscattered data. The radar image is obtained by the Fourier transform of the recovered data matrix. The resulting radar images satisfy the continuity assumption of the target. Their approach, where the missing data problem is solved by a preprocessing step involving  data reconstruction with appropriate dictionaries, can be considered as a matrix completion procedure.

 Matrix completion (MC) methods are based on low-rank matrix approaches and  have been applied to many missing data case problems in different domains \cite{survey_mc,survey_mc2}.
 In fact, the obtained ISAR matrix also exhibits low‐rank property. Thus, MC methods can be used to solve the missing case problem under certain conditions by using the low-rank constraints in ISAR imaging.
Singular value thresholding (SVT)\cite{SVT} algorithm can be easily applied to handle a random sampling model. The processing can be performed directly in matrix form, thus avoiding the vectorization operation of traditional CS algorithms. In \cite{RPCA}, the MC problem has been converted into the robust principal component analysis (RPCA) problem. 
Fast MC methods have also been proposed in image processing domain and they were recently applied to the radar signal reconstruction problems in the work of \cite{R_MC}.  Go Decomposition (GoDec) provides the low rank component in a SVD-free way by using bilateral random projections (BRP). In \cite{R_MC}, it was shown that nuclear norm minimization (NNM) method has a faster and superior performance for radar signal reconstruction in  missing data case compared to other conventional methods for both simulated and real datasets. There is also  Inexact Augmented Lagrangian Multipliers (IALM) method  which also uses low rank property of the data matrix. Apart from the NNM, IALM can be used directly on complex data. Thus, it does not require any preprocessing such as separation to perform completion on missing matrices. In fact, the aforementioned MC methods generally have some limitations, such as excessive sampling case on the random sampling pattern. 

In this work we propose to complete the missing entries of the data matrix using deep image prior (DIP). DIP has already been used for the completion of GPR images in \cite{DK_DIP} to enhance detection rates of the buried targets. However in \cite{DK_DIP}, GPR data is collected in time domain, thus cascading of the measurements gives directly real valued raw GPR image (B-scans) on which detection process is implemented. In our work radar data are collected in polar coordinates and  have to be converted to cartesian frequency coordinates after the motion compensation is applied. Finally 2D Fourier transform is used as imaging method to provide 2D radar images thus unlike \cite{DK_DIP}, first 2D complex frequency domain data should be completed and then an imaging algorithm should be applied.
 
DIP employs a fixed weight deep network as hand crafted prior and it does not require any training. As it works on a single image iteratively, it provides a huge advantage compared to pre-trained network models. DIP uses Deep Neural Network (DNN) to perform different tasks such as denoising, inpainting and super resolution on real images. 
 Motivated by the success of DIP for GPR image completion \cite{DK_DIP}, we adapt the idea to work on complex data domain. Same model is employed to sequentially complete the real and imaginary parts of the radar echo. The loss function utilizes the known entries to calculate the mean squared error (MSE) between the reference and completed coefficients at each iteration thus, network parameters are updated with  calculated loss. Once the raw data matrix is completed, the ISAR image can be obtained by 2D IFFT or other imaging methods such as 2D MUSIC or 2D AR modelling .
To the best of our knowledge this is the first study which aims to complete raw ISAR data by using DNN without any training process in data domain.

The rest of the paper is organized as follows. Section II illustrates the signal model for ISAR imaging. In Section III, general missing data recovery formulation is introduced and extended for the complex signals. In Section IV, the missing data completion method using DIP is proposed. The comparisons of methods for both simulated and real datasets are presented and discussed in Section V and finally the paper is concluded in Section VI.

\section{Signal Model}

\begin{figure}[t!]
\begin{center}
\includegraphics[width=0.17\textwidth]{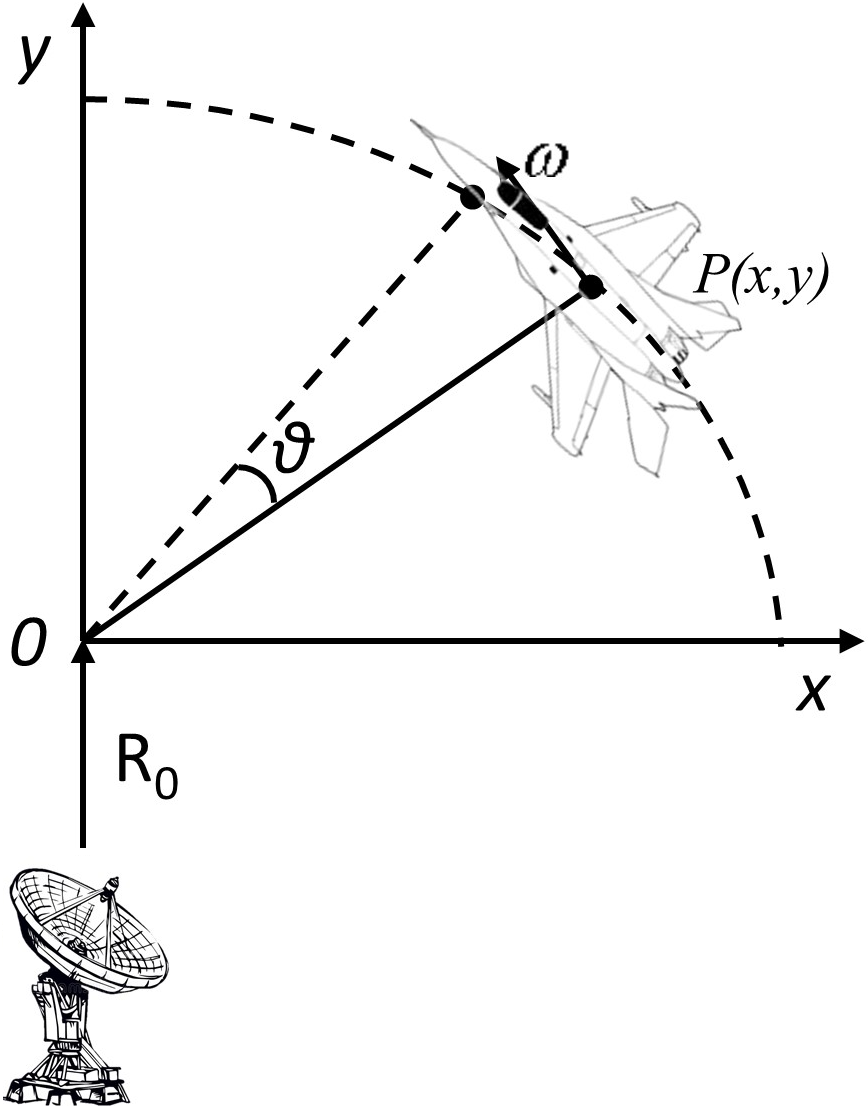}
\caption{ISAR geometry representation.}
\label{fig:1}       
\end{center}
\end{figure}

Let us consider the ISAR imaging scenario given in Fig. \ref{fig:1}. Here the translational motion is assumed to be compensated, so we only have a relative rotational motion between radar and the target. According to far field assumption, the instantaneous distance between the point P(x,y) and the target can be written as

\begin{equation}
\mathrm{R(t)} \approx \mathrm{R_{0}} + xsin({\omega}t) + ycos({\omega}t) 
\label{target_range}
\end{equation}
where $R_{0}$ is the range from radar to the origin and $\omega$ is the angular rotational speed thus $\omega$$t$ corresponds to rotation angle at time $t$. For stepped frequency radar, radar transmits N pulses with $f_{0}$ as the initial frequency and ${\Delta}f$ as the frequency step size.
As a same way of frequency, angle between radar to target is also changed with angular step of radar. 
So $f_{n} = f_{0} + (n-1){\Delta}f$ and  $\theta_{m} = (m-1){\Delta}{\theta}$ correspond to  $nth$ frequency and $mth$ angle respectively. By using this notation, the received signal can be derived for $mth$ burst of $nth$ pulse as

\begingroup
\scriptsize
\begin{equation}
\begin{split}
Y(m,n) &= \sum_{k=1}^K \alpha_k \exp\left(-j \frac{4 \pi f_n R_k}{c} \right) \\
      &= \sum_{k=1}^K \alpha_k \exp\left(-j \frac{4 \pi f_n}{c} \left(R_0 + x_k \sin\theta_m + y_k \cos\theta_m \right) \right)
\end{split}
\label{recieved_signal}
\end{equation}
\endgroup

Here K is the total number of scatterers, $c$ is speed of light, $R_{k}$
corresponds to the distance from the $kth$ scatterer to the radar, $x_{k}$ and $y_{k}$ are the horizontal and vertical distances to $kth$ scatterer respectively. After omitting the constant phase term $\mathrm{exp}\left(-j\frac{4{\pi}f_{n}{R_0}}{c}\right)$,  under the small rotational angle condition $sin\theta_{m} \approx \theta_{m}$, $cos\theta_{m}$  $\approx 1$, equation (\ref{recieved_signal}) can rewritten as 

\begingroup
\scriptsize
\begin{equation}
Y(m,n) = \sum_{k=1}^K \alpha_k \exp\left(-j \frac{4 \pi f_n x_k \theta_m}{c} \right) \exp\left(-j \frac{4 \pi f_n y_k}{c} \right)
\label{simplified_recieved}
\end{equation}
\endgroup

 The coordinates of  $kth$ scatterer $(x_{k},y_{k})$ can be discretized as $x_{k} = p{\Delta}_{x}$ and $y_{k} = q{\Delta}_{y}$ where ${\Delta}$y and ${\Delta}$x denote range and cross-range resolution of radar respectively while $p = 1,2,...,M$ and $q = 1,2,...,N$. Resolutions can be defined as ${\Delta}_{y} = \frac{C}{2N{{\Delta}f}}$ and ${\Delta}_{x} = \frac{C}{2M{{\Delta}\theta}}$. Substituting these on (\ref{simplified_recieved}) the measurement data can be finally expressed as 
 
\begingroup
\begin{multline}
\scriptsize
Y(m,n) = \sum_{p=1}^M \sum_{q=1}^N \alpha_{p,q} 
\exp\left(-\frac{j 2 \pi p (m-1)}{c}\right) \\
\times \exp\left(-\frac{j 2 \pi q (n-1)}{c}\right)
\label{last_recieved}
\end{multline}
\endgroup

As a last step, 2D Fourier transform is applied to \ref{last_recieved} to obtain the ISAR image.

\section{Data Recovery For Complex ISAR Data With Missing Entries}
\label{sec:missing_data_rcv_bscans}

 ISAR data matrix can be depicted as $Y \in \mathbb{C}^{m \times n}$  and $M \in \mathbb{C}^{m \times n}$ denotes the ISAR data matrix with missing entries where $m$ and $n$ represent the dimensions, namely total frequency step and scanned angles. Complex matrix $M$ can be defined as $M = M_{R} + jM_{I}$ where $M_{R}$ and $M_{I}$ are real and imaginary coefficient of the ISAR data. Since, $M$ is a complex matrix and sampling operator is defined as $P_{\Omega} : \mathbb{R}^{m \times n} \xrightarrow{} \mathbb{R}^{m \times n}$, recovery of the missing samples can be formulated real and imaginary parts separately as follows

\begin{subequations}
\begin{equation}
\big[P_{\Omega}(M_{R})\big]_{i,j} = 
\begin{cases}
    M_{R_{i,j}}(i,j) \in   \Omega\\
    0, \quad \text{otherwise}
\end{cases}\label{eq:rcv_missingData}
\end{equation}

\begin{equation}
\big[P_{\Omega}(M_{I})\big]_{i,j} = 
\begin{cases}
    M_{I_{i,j}}(i,j) \in   \Omega\\
    0, \quad \text{otherwise}
\end{cases}\label{eq:rcv_missingDataImag}
\end{equation}
\end{subequations}

where $\Omega$ denotes the known entries, and $P_{\Omega}$ represents a sampling operator in the observed region $\Omega$. 

If the missing data matrix $M$ is low rank and its singular values are spread enough, the missing entries of the data matrix can be recovered by matrix completion theory \cite{crb_algorithm} which uses the low rank property of the partially observed input matrix. The  optimization problem can be formulated as

\begin{subequations}
\begin{equation}
\min_{Z_R} \; \mathrm{rank}(Z_R) \quad \text{s.t.} \quad \left\lVert P_{\Omega}(Z_R - M_{R}) \right\rVert_{\mathrm{F}} \leq \delta
\label{eq:optim_prb_real}
\end{equation}
\begin{equation}
\min_{Z_I} \; \mathrm{rank}(Z_I) \quad \text{s.t.} \quad \left\lVert P_{\Omega}(Z_I - M_{I}) \right\rVert_{\mathrm{F}} \leq \delta
\label{eq:optim_prb_imag}
\end{equation}
\end{subequations}

where $Z_R$ and $Z_I$ are the randomly initialized unknown variable matrices, $\left\lVert \cdot \right\rVert_{\mathrm{F}}$ denotes the Frobenius norm of the difference matrix, $\delta$ is a tolerance parameter that limits the error. Since the optimization problems in \eqref{eq:optim_prb_real}  and \eqref{eq:optim_prb_imag} are NP-hard, they should be converted to the following convex optimization problem as 

\begin{subequations}
\begin{equation}
\operatornamewithlimits{min}_{Z_R}\left\lVert Z_{R} \right\rVert_* \quad \text{s.t.} \quad \left\lVert P_{\Omega}(Z_R-M_R) \right\rVert_{\mathrm{F}} \leq \delta \label{eq:convex_opt_real}
\end{equation}
\begin{equation}
\operatornamewithlimits{min}_{Z_I}\left\lVert Z_I \right\rVert_* \quad \text{s.t.} \quad \left\lVert P_{\Omega}(Z_I-M_I) \right\rVert_{\mathrm{F}} \leq \delta \label{eq:convex_opt_imag}
\end{equation}
\end{subequations}

Here $\left\lVert \cdot \right\rVert_*$ denotes the nuclear norm operation. The optimization problem formulated in \eqref{eq:convex_opt_real}, \eqref{eq:convex_opt_imag} can be solved by matrix completion methods \cite{GoDec, FAN201834}.

As given in \cite{R_MC}, a pre-transformation is required  for the column-wise case to recover the input ISAR data matrix with missing data since at least one observation for each row and column is necessary according to matrix completion theory ~\cite{crb_algorithm}. More details about the pre-transformation step can be found in ~\cite{R_MC}.

\begin{figure*}[t!]
\begin{center}
\includegraphics[width=1\textwidth]{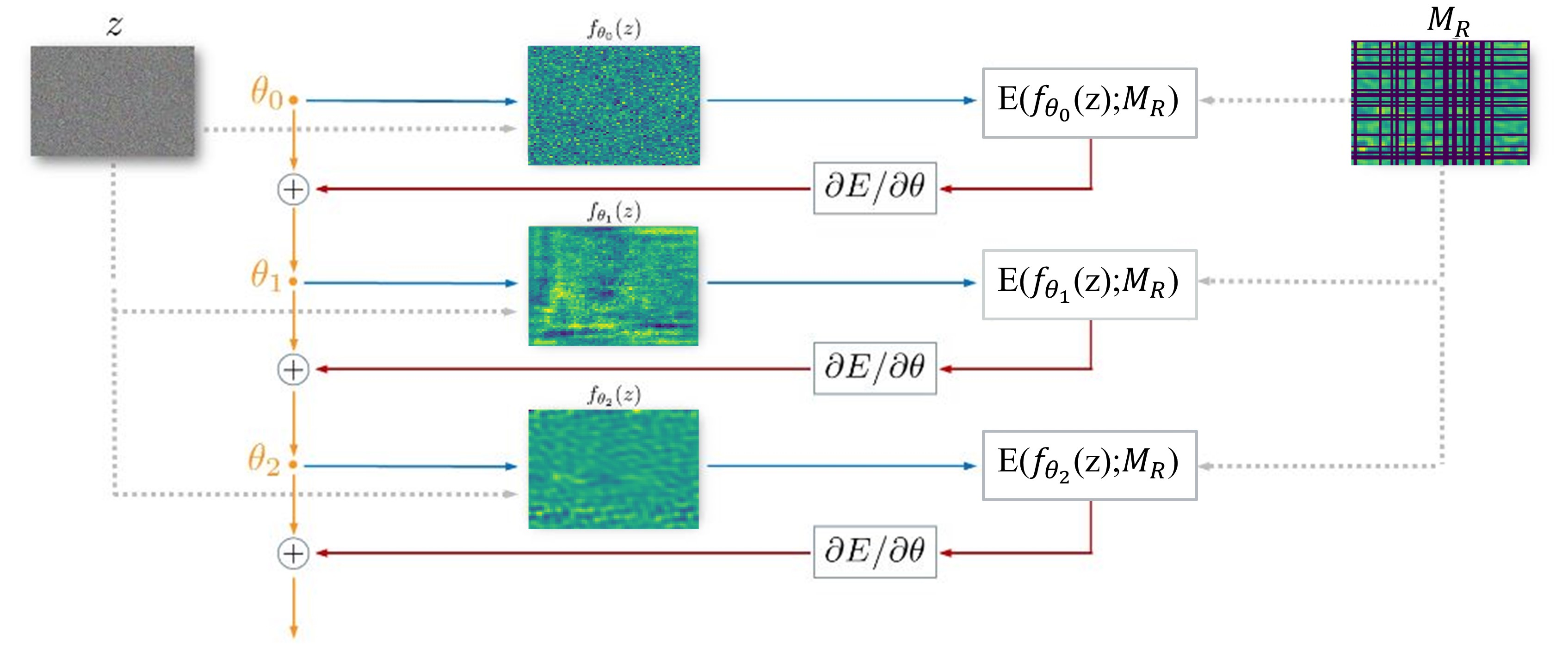}
\caption{Missing Data recovery steps of DIP method. }
\label{fig:2}       
\end{center}
\end{figure*}

\section{Proposed DIP-Based ISAR Imaging Method} \label{sec:proposedmethod}

Matrix completion algorithms aim to recover the original image from the image with missing entries. In DIP, the recovery task can be formulated  as an optimization problem as

\begin{equation}\label{eq:energyMinimization}
    Z_R^* = \min_{Z_R} \left[ E(Z_R; M_R) + R(Z_R) \right]
\end{equation}
where $E(Z_R,M_R)$ represents data term
and $R(Z_R)$ is the regularization term. The data term $E(Z_R,M_R)$ is usually easy to design for a wide range of problems, such as super-resolution, denoising, inpainting, while image prior $R(Z_R)$ is  captured with a ConvNet.
For the  data recovery case, energy function can be given as \eqref{eq:energyCorpImg}

\begin{equation}\label{eq:energyCorpImg}
    E(Z_R; M_R) = \Vert{P_{\Omega}}(Z_R - M_R)\Vert^2
\end{equation}

where $(Z_R-M_R)$ represents the difference between reconstructed signal and original signal with missing entries. As it is seen in \eqref{eq:energyCorpImg} missing entries are neglected in energy calculation. Thus, if energy is calculated directly by pixel values, it would never change after initialization so that \eqref{mapped_signal} is utilized.

Deep generators are focused on extracting high level priors to handle the best regularization term, so they require a high amount of training data to extract generic priors from the image set. Instead, DIP is focused on hand crafted priors to replace explicit regularization term $R(Z_R)$ with implicit mapping function. The aim is to reconstruct image by using single corrupted image.  The image is recovered as
\begin{equation}
    Z_R^* = f_{\theta^*}(z).
\label{mapped_signal}
\end{equation}

Here $f$ is a deep neural network structure with parameters ${\theta}$ initialized randomly and $z$ is a fixed input generally chosen as white  noise. The network parameters are updated with respect to \eqref{eq:minimizer} iteratively.
\begin{equation}\label{eq:minimizer}
    \theta^* = \arg\min_{\theta} E(f_\theta(Z_R; M_R))
\end{equation}

The gradient descent can be used to optimize the modified energy function in \eqref{eq:minimizer}, for an image 
$Z_R \in \mathbb{R}^{1 \times H \times W}$, code vector   $z \in \mathbb{R}^{C' \times H' \times W'}$. Here, $H$ and $H'$ are used for height, $W$ and $W'$ for width and $C'$ for code tensor, respectively. Complete iteration scheme of DIP method can be seen in Fig. ~\ref{fig:2}.

In summary, DIP steps can be given as:
\begin{itemize}
\item Initialize the neural network's parameters randomly. 

\item Apply the random noise as input to the neural network to generate an output image. 

\item Compare the output image with the input image using a loss function (e.g., pixel-wise difference, perceptual loss). 

\item Backpropagate the gradient of the loss with respect to the network's parameters and update the parameters using an optimization algorithm (e.g., gradient descent). 

\item Repeat steps 2-4 for multiple iterations or until convergence. 
\end{itemize}

\begin{figure}[h!]
    \centering
    \includegraphics[width = 0.9\linewidth]{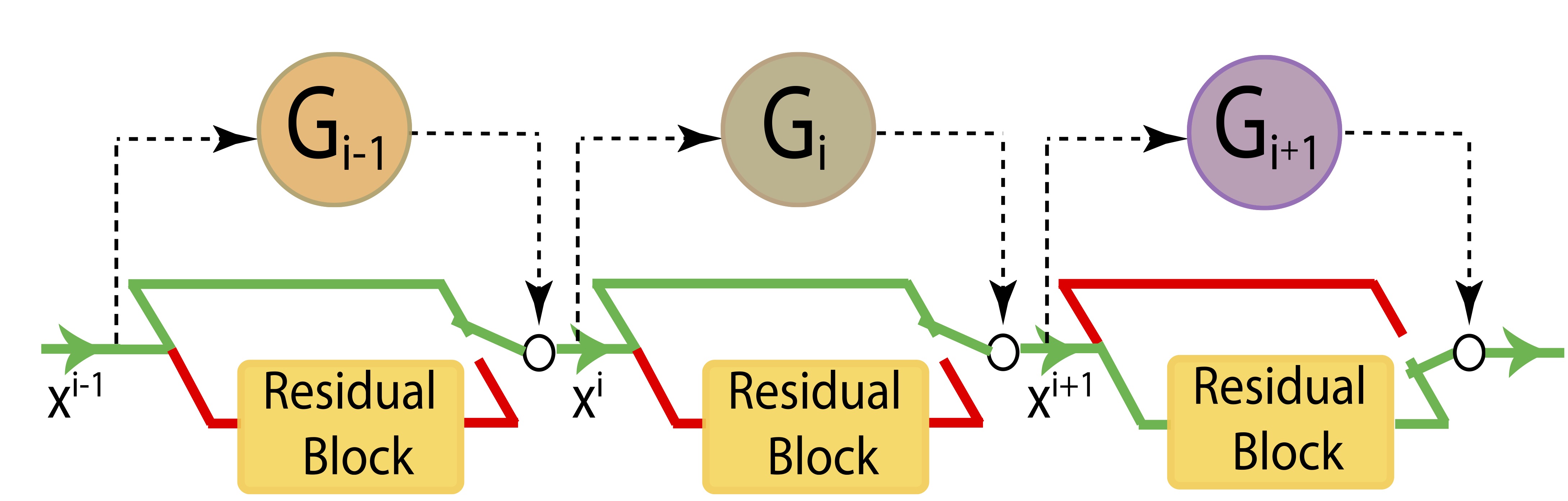}
    \caption{SkipNet Architecture that is embedded into DIP model.}
    \label{fig:3}  
\end{figure}

\begin{figure*}[h!]
    \centering
    \includegraphics[width = 0.94\linewidth]{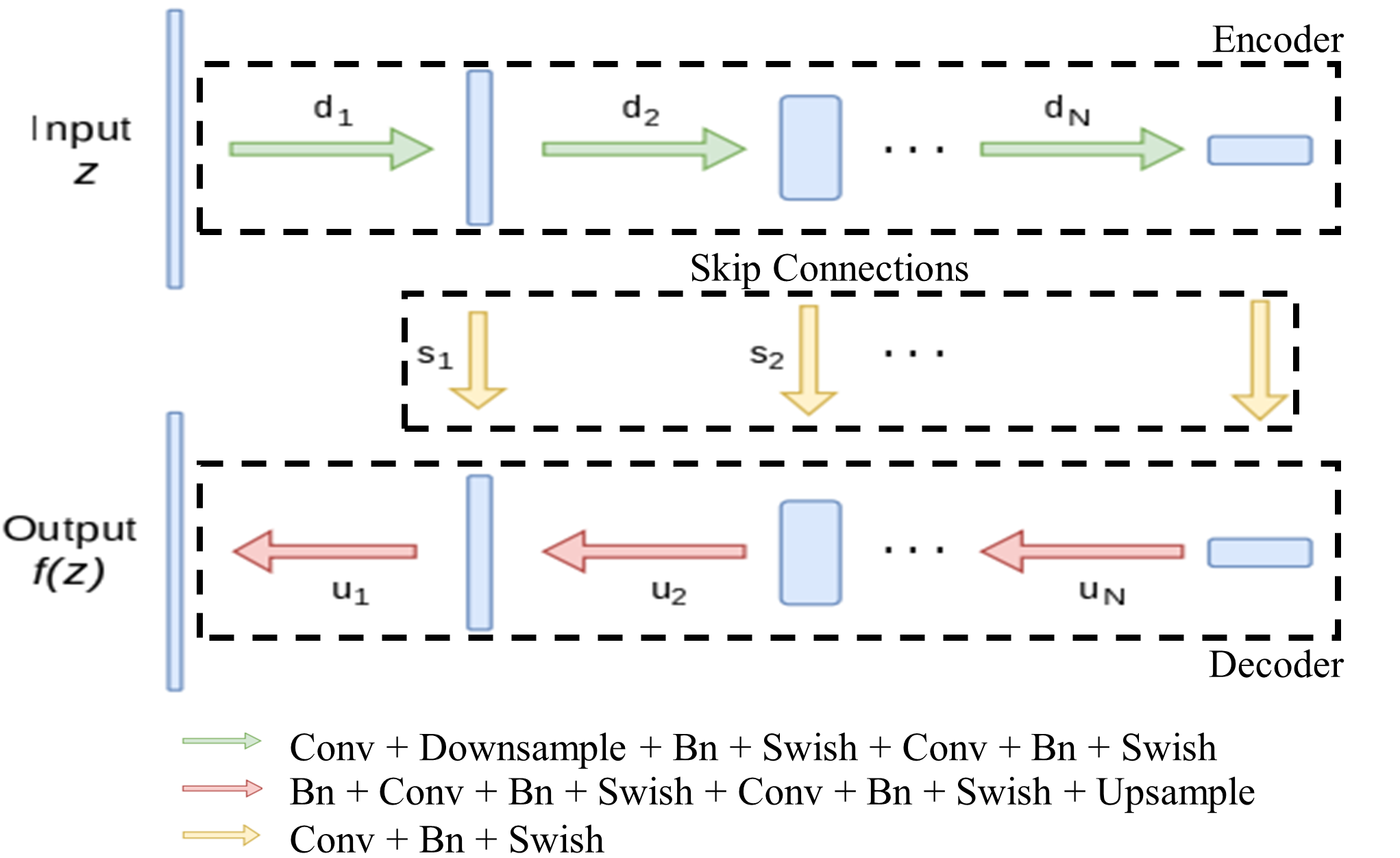}
    \caption{Layers of the DIP network with skip connections.}
    \label{fig:4}
\vspace{2mm}    
\end{figure*}

DIP can provide, SkipNet, Unet and ResNet as model architecture. In this study SkipNet is used \cite{skipnet} as our model which provides skip connection between layers, thus some layers can be skipped or inserted as seen in Fig.~\ref{fig:3}. This feature is controlled by small gating networks which map the output of the previous layer or group of layers to binary decision to skip or run next layer or group of layers. Skip connection's can be expressed as
\begin{equation}\label{skipnet_eq}
    x^{i+1} = G^i(x^i)F^i(X^i) + (1-G^i(x^i))x^i,
\end{equation}
where $x^i$ is the input, $F^i(x^i)$ is the output of the $i^{th}$ layer or layers and $G^i(x^i) \in \{0,1\}$ is the gating function of the $i^{th}$ layer.

As it can be seen in Fig.~\ref{fig:4}, there are skip connections between encoder and decoder blocks thus, the skip operation that is shown in Fig.~\ref{fig:3} are applied to DIP model.

The accuracy of the model can be improved with model depth, however the inference latency and complexity of the optimization function would be dramatically increased. To avoid this trade-off, the skip connection provides adaptive depth to the model by varying model depth in accordance to the quality of the input image. Therefore, it reduces to computation cost for reasonable input while it improves the accuracy of the model. The details of the skip connection effect on complexity of the minimization problem can be found in \cite{loss_of_nets}.

\begin{figure*}[h!]
\begin{center}
\includegraphics[width=0.9\textwidth]{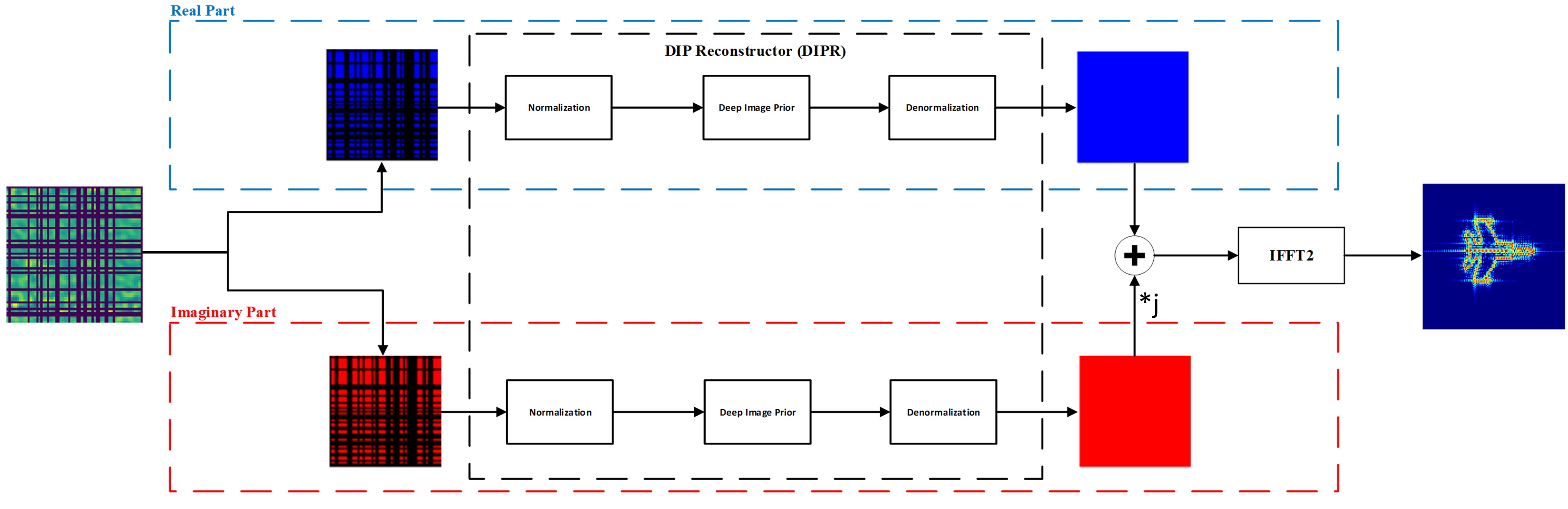}
\caption{Flow diagram of Proposed Method.}
\label{fig:5}  
\end{center}
\end{figure*}

Since DIP is not designed to work with complex valued data, raw ISAR data is separated into its real and imaginary parts as it is seen on Fig.~\ref{fig:5}.
Separating ISAR data into its real and imaginary parts to feed coefficients into the DIP is formulated as below.
\begin{equation}\label{DIPR}
\mathrm{Y}_{\mathrm{R}} = ||\Re{(M)}||_\mathrm{{DIPR}} + j||\Im{(M)}||_\mathrm{{DIPR}}
\end{equation}
where $||.||_{\mathrm{DIPR}}$ denotes to Deep Image Prior based reconstruction which also added normalization and denormalization steps to classical DIP, $\Re$ and $\Im$ are real and imaginary part of incomplete matrix, $M$ is equal to incomplete ISAR raw data and $\mathrm{Y_{R}}$ represents to reconstructed complex ISAR data. By using the (\ref{DIPR}), the equations that are between \eqref{eq:energyMinimization} to \eqref{eq:energyCorpImg} are applied to real and imaginary part separately. Since ISAR image is created by applying 2-D Fourier transform to raw data, the reconstructed signal has to be in exact same level with input signal. As it is seen in Fig.~\ref{fig:5}, in the proposed method, normalization and denormalization steps are added to overcome scale loss.

As a default structure, SkipNet based DIP uses a model depth of four, and skip connections are not reinforced with convolutional layers for image reconstruction tasks. Experimental results show that increasing the channel size improves the signal reconstruction performance till some point, then reduces it.~To get better reconstruction results, convolutional layers are added to skip connections.~Thus, improved results are obtained with the same channel size for convolutional layers.~The obtained results may be handled by fine-tuning the channel size, however adding convolutional layers to skip connections improves the complexity of skip connections and it is beneficial for signal reconstruction performance of the proposed DIP. As it is seen in Fig. \ref{fig:5}, inputs are normalized between 0 and 1 and best results are obtained with Swish thus, it is used as activation function. Adam optimizer is used with learning rate as 1e-3 and MSE as loss function. As it is previously noted, model depth is set as six and channel sizes are set as 256,128,64,64,128,256 for both encoder and decoder blocks. 5x5 kernels are used in convolution layers.

\section{Experimental Results}

In this section, both visual and quantitative performance evaluations are presented to compare DIP based signal reconstruction with different imaging methods. Both real and simulated ISAR images are compared by using three different scenarios as

\begin{itemize}
    \item Pixel-wise missing data case
    \item Column-wise missing data case
    \item Compressed missing data case
\end{itemize}

 Random pixel-wise (point-wise) loss, column-wise (azimuth) loss and compressed data scenarios are investigated for three different missing rates of $30\%$, $50\%$, and $70\%$, respectively. In the compressed case, restricted samples of raw data are used so it can also be taken as a missing case. Three different performance comparison metrics are used for quantitative evaluations. The formulation of quantitative performance metrics RMSE, Correlation and Contrast are given in below, respectively. 

\begin{equation}
\mathrm{RMSE} = \frac{{\left | \left| \mathrm{I} - \hat{\mathrm{I}} \right |\right|}_{\mathrm{F}}}{{\left |\left|\mathrm{I}\right|\right|}_{\mathrm{F}}}
\label{rmse_eq}
\end{equation}

RMSE score of the reconstructed signal is calculated by using \eqref{rmse_eq}. $F$ represents the Frobenius norm of matrix, $\mathrm{I}$ and $\mathrm{\hat{I}}$ represent the original and reconstructed ISAR images, respectively. ISAR images are generated by taking 2-D Fourier transform of backscattered field $Y$ and reconstructed backscattered field $Y_R$. RMSE computes the difference between each pixel in the reconstructed image and the corresponding pixel in the original image. A lower RMSE score indicates a better reconstruction performance. The evaluation result is also normalized using the Frobenius norm of the original image.

\begin{equation}
\mathrm{Corr(I,\hat{I})} = \frac{1}{N-1}{\sum_{i=1}^{N}}{\left( {\frac{I_{i} - \mu_{I}}{\sigma_{I}}} \right)}{\left( {\frac{\hat{I}_{i} - \mu_{\hat{I}}}{\sigma_{\hat{I}}}} \right)}
\label{correlation_eq}
\end{equation}

The correlation is computed by \eqref{correlation_eq} where $\mathrm{I}$ and $\mathrm{\hat{I}}$ correspond to the vectorized reference image and the reconstructed image, respectively. $\mu$ is the mean and $\sigma$ is the standard deviation of given images. Correlation measures the coherence between the original and reconstructed images, and a higher correlation indicates better imaging performance.

\begin{equation}
\mathrm{IC} = \frac{1}{MN} \sum \frac{\left( \sqrt{ \left( \hat{I}^2 - \mu_{\hat{I}^2} \right)^2 } \right)}{\mu_{\hat{I}^2}}
\label{contrast_eq}
\end{equation}

Image contrast (IC) is one of the selected metrics for evaluating performance and it is frequently used in this field. In \eqref{contrast_eq}, $\mathrm{\hat{I}}$ corresponds to the reconstructed image and $\mathrm{\mu_{\hat{I}^2}}$ corresponds to the mean value of the $\mathrm{\hat{I}^2}$. $M$ and $N$ represent the image row and column sizes, respectively. The calculation of IC is performed using only the reconstructed signal, so it is not sufficient to evaluate the reconstruction performance on its own. However, if the RMSE and correlation values are similar to each other, IC can provide some additional insight about the performance comparison.

\subsection{Simulated Dataset Results}

\begin{figure}[!h]

\begin{center}
\includegraphics[width=.45\textwidth]{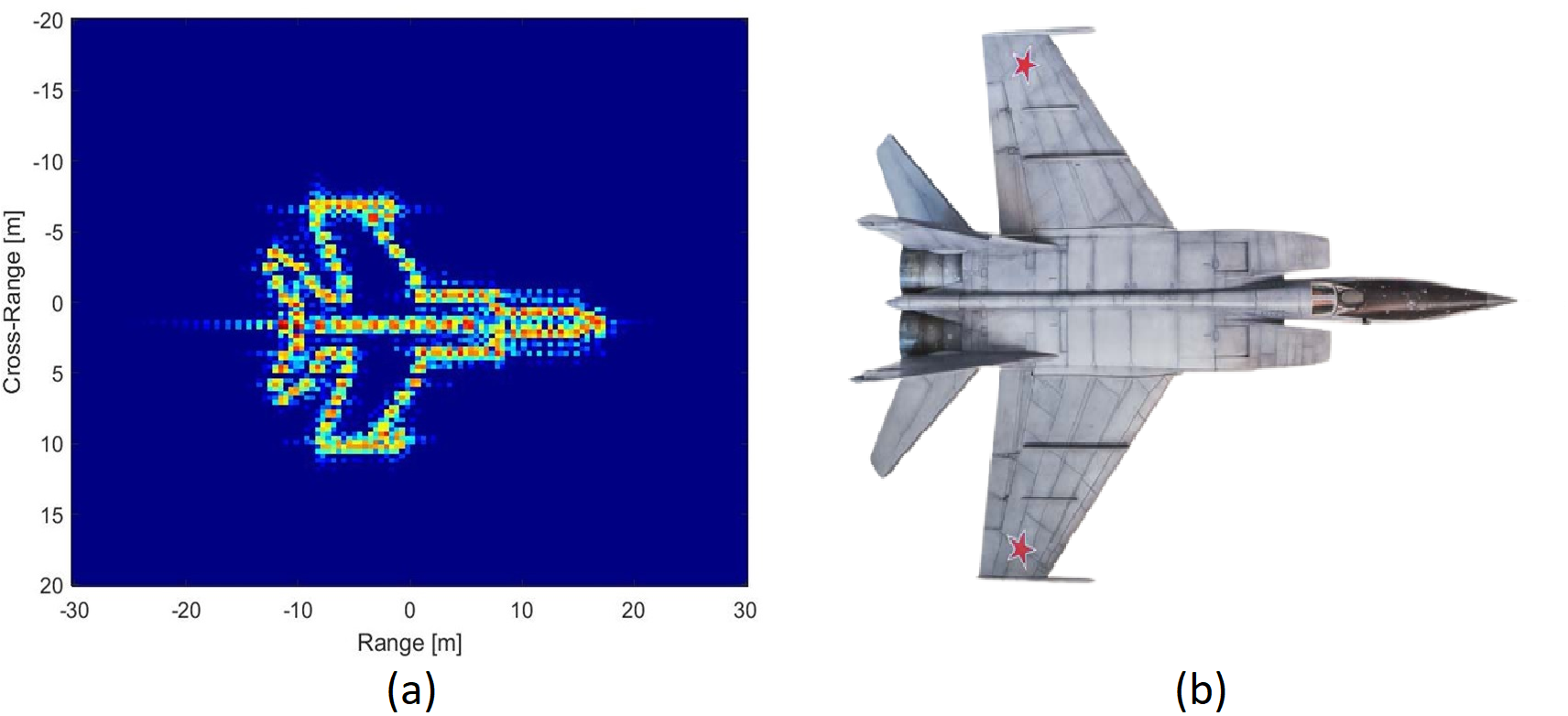}
\caption{(a) ISAR imaging result of simulated Mig-25 warplane with full data by using RD algorithm. ISAR image is normalized to the maximum amplitude and shown on a top 20 dB; (b) Photo of Mig-25 warplane.}
\label{fig:6}
\end{center}
\end{figure}

\begin{figure}[!h]

\begin{center}
\includegraphics[width=.49\textwidth]{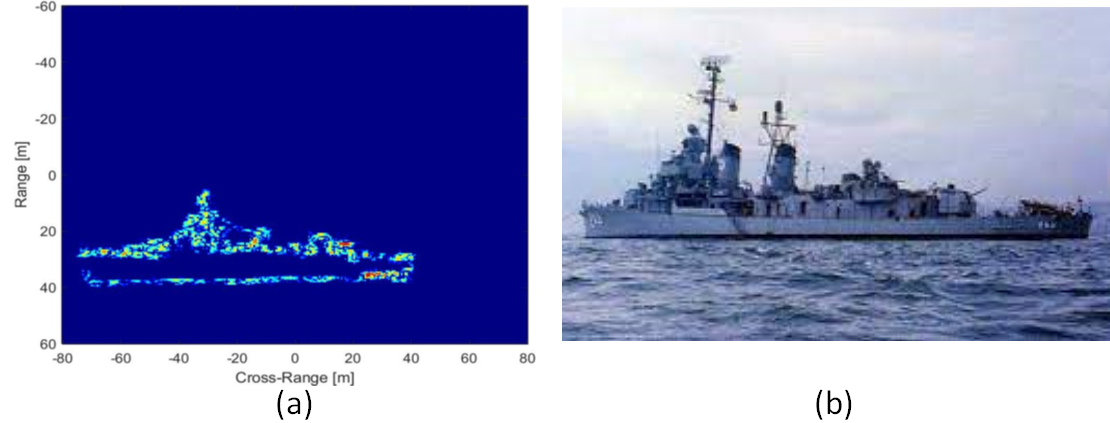}
\caption{(a) ISAR imaging result of simulated USS Fletcher warship with full data by using RD algorithm. ISAR image is normalized to the maximum amplitude and shown on a top 20 dB; (b) Photo of USS Fletcher warship.}
\label{fig:7}       
\end{center}
\end{figure}

Simulated Mig-25 warplane and USS Fletcher warship are used to evaluate model performance for 3 different missing cases. The ISAR images of the simulated data and their real images are presented in Fig. \ref{fig:6} and Fig. \ref{fig:7}, respectively. The brief technical details of the simulated datasets can be given as follows; simulated Mig-25 data has 8 GHz center frequency and 531 MHz bandwidths and view angle is 3.67°. Simulated USS Fletcher has 9 GHz center frequency and 125 MHz bandwidth and view angle is 2.36°  \cite{caner_hoca}. 
Both of the simulated data are tested under 3 different missing cases such as pixel-wise, column-wise (azimuth) missing case and compressed case. The results are evaluated both visually and quantitatively.

\begin{figure*}[!h]
\begin{center}
\includegraphics[width=0.67\textwidth]{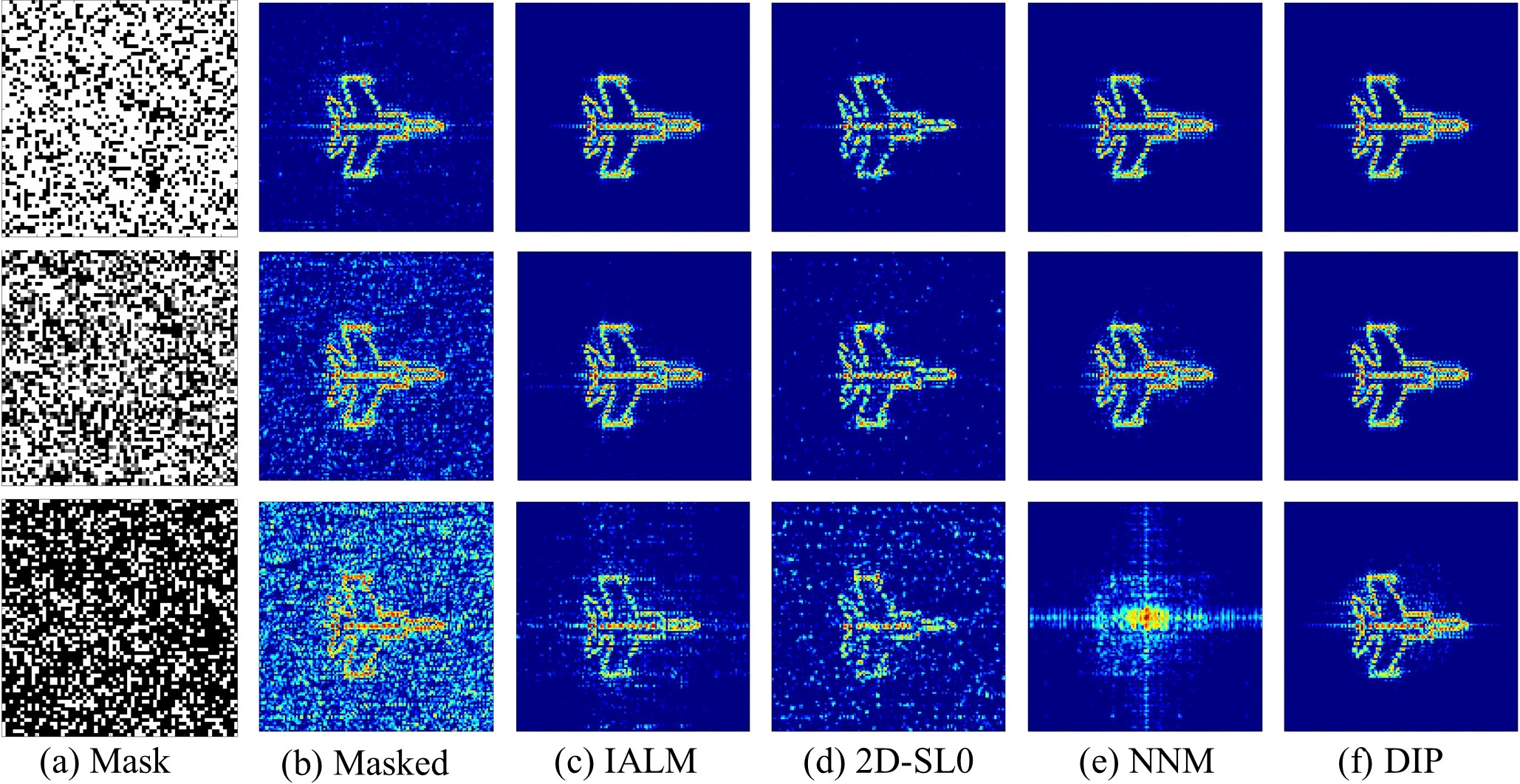}
\caption{Visual comparison of ISAR imaging completion methods for simulated Mig-25. The pixel-wise missing rates (\%) are: 30, 50, and 70 from top to bottom. (a) Applied masks, (b) the corrupted ISAR image and the reconstruction results for (c) IALM, (d) 2D-SL0, (e) NNM, and (f) proposed DIP. Images are normalized to their maximum and top 20 dB are shown.}
\label{fig:8}       
\end{center}
\end{figure*}

\begin{table*}[h]
\caption{Quantitative comparison results of ISAR imaging methods for simulated Mig-25 data in pixel-wise missing case.} 
\centering 
  \resizebox{13.5 cm}{!}{%
\begin{tabular}{c c c c c c c c c c c c} 
\toprule
\multirow{ 2}{*}{\textbf{Method}} &\multicolumn{3}{c}{\textbf{RMSE}} & & \multicolumn{3}{c}{\textbf{Correlation}} & &  \multicolumn{3}{c}{\textbf{Contrast}}\\
\cmidrule{2-4} 
\cmidrule{6-8}
\cmidrule{10-12}

  & L = 30\% & L = 50\% & L = 70\% & & L = 30\% & L = 50\% & L = 70\% & & L = 30\% & L = 50\% & L = 70\% \\ \cmidrule(r){1-12}
IALM	&	0.0851	&	0.2950	&	0.5817	&&	0.9957	&	0.9612	&	0.8430	&&	1.644	&	1.4949	&	1.1828	\\
2D-SL0	&	0.4919	&	0.5706 &	0.8431	&&	0.8417	&	0.8061	&	0.6151	&&	1.4504	&	1.3099	&	1.0430	\\
NNM	&	0.1175	&	0.3316	&	0.8157	&&	0.9923	&	0.9487	&	0.4747	&&	1.6264	&	1.4897	&	1.4851	\\
Proposed DIP	&	\textbf{0.0407}	&	\textbf{0.0832}	&	\textbf{0.3139}	&&	\textbf{0.9989}	&	\textbf{0.9958}	&	\textbf{0.9447}	&&	\textbf{1.6602}	&	\textbf{1.6555}	&	\textbf{1.5646}	\\
\hline
\end{tabular}
}
\label{tab:1} 
\end{table*}

\begin{figure*}[!h]
\begin{center}
\includegraphics[width=.67\textwidth]{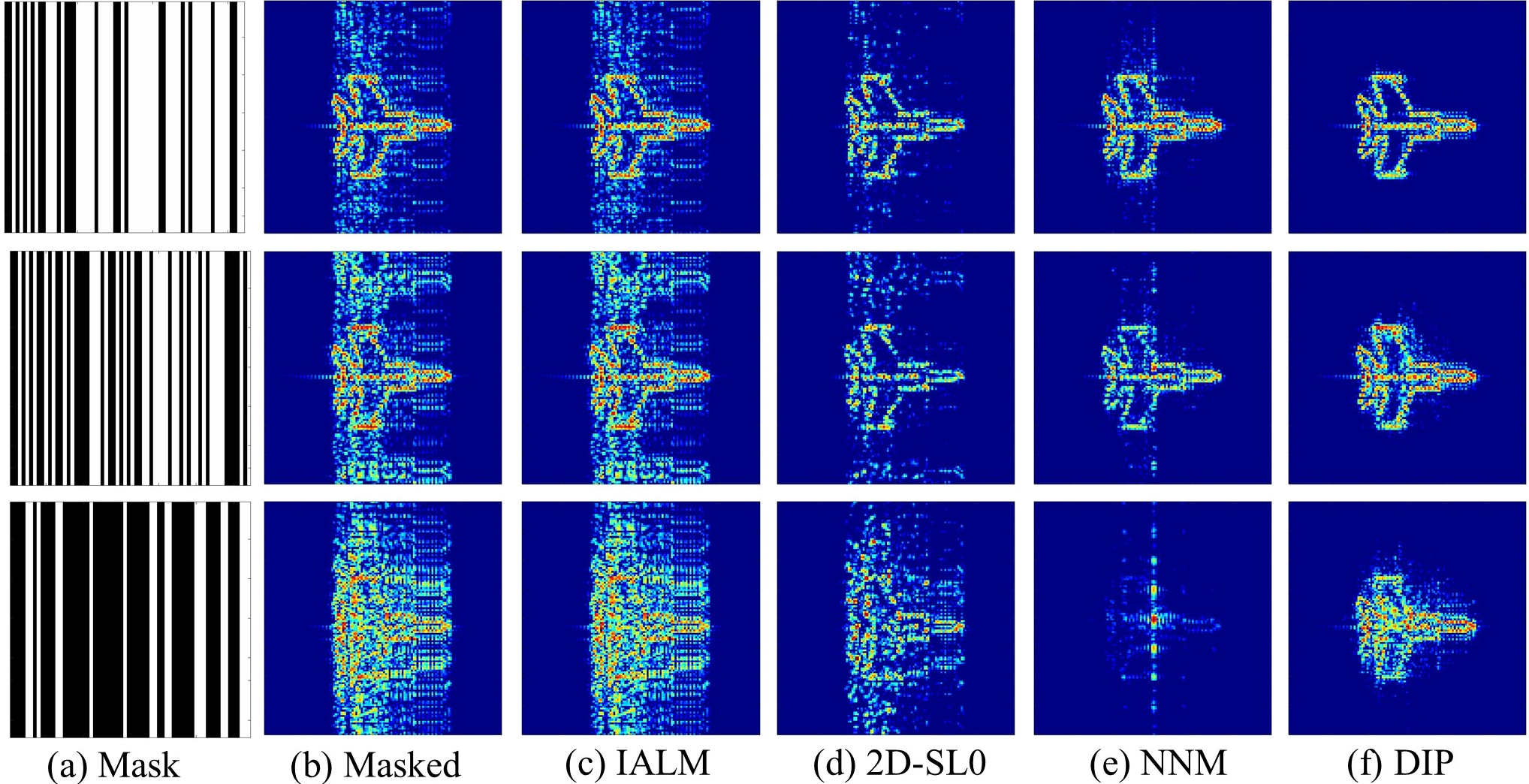}
\caption{Visual comparison of ISAR imaging completion methods for simulated Mig-25. The missing column rates (\%) are: 30, 50, and 70 from top to bottom. (a) Applied masks, (b) the corrupted ISAR image and the reconstruction results for (c) IALM, (d) 2D-SL0, (e) NNM, and (f) proposed DIP. Images are normalized to their maximum and top 20 dB are shown.}
\label{fig:9}       
\end{center}
\end{figure*}

\begin{table*}[!h]
\caption{Quantitative comparison results of ISAR imaging methods for simulated Mig-25 data in column-wise missing case.} 
\centering 
  \resizebox{13.5 cm}{!}{%
\begin{tabular}{c c c c c c c c c c c c} 
\toprule
\multirow{ 2}{*}{\textbf{Method}} &\multicolumn{3}{c}{\textbf{RMSE}} & & \multicolumn{3}{c}{\textbf{Correlation}} & &  \multicolumn{3}{c}{\textbf{Contrast}}\\
\cmidrule{2-4} 
\cmidrule{6-8}
\cmidrule{10-12}

  & L = 30\% & L = 50\% & L = 70\% & & L = 30\% & L = 50\% & L = 70\% & & L = 30\% & L = 50\% & L = 70\% \\ \cmidrule(r){1-12}
IALM	&	0.6409	&	0.9007	&	1.1477	&&	0.8590	&	0.7632	&	0.6376	&&	1.3673	&	1.3259	&	1.2674	\\
2D-SL0	&	0.5374	&	0.6303	&	0.8427	&&	0.8116	&	0.7296	&	0.6098	&&	1.4772	&	1.3971	&	1.3653	\\
NNM	&	0.3586	&	0.4505	&	0.9184	&&	0.9352	&	0.8682	&	0.3270	&&	1.5492	&	1.4781	&	1.4146	\\
Proposed DIP	&	\textbf{0.18427}	&	\textbf{0.35286}	&	\textbf{0.5334}	&&	\textbf{0.9788}	&	\textbf{0.9378}	&	\textbf{0.8438}	&&	\textbf{1.6640}	&	\textbf{1.6125}	&	\textbf{1.5566}	\\

\hline
\end{tabular}
}
\label{tab:2} 
\end{table*}

Pixel-wise missing case masks are presented in Fig. \ref{fig:8}(a). The results of RD imaging are given in Fig. \ref{fig:8}(b) which shows the degradation in the performance from top to bottom as the total number of missing samples increases. Additionally, the presence of side lobe effects and high levels of artifacts can be observed in the RD imaging results. The imaging results of the comparison and the proposed methods are presented in Fig. \ref{fig:8}(c)--(f), respectively. The first observation is that all imaging methods showed better performance compared to the RD imaging. Similar to the results of RD imaging, the  imaging results present a decrease in performance from top to bottom with the increasing missing ratio. 

Our first experiment is 30\% missing ratio and this is the simplest case. All imaging methods, except for 2D-SL0 demonstrated similar performances. There are some undesired attenuation on the target peak for the 2D-SL0. It is hard to notice the difference between the remaining methods by only visual inspection.

In the second case, 50\% missing ratio is used for performance comparison. The artifacts are significantly increased in the RD imaging result. For the reconstruction results of the comparison methods, IALM and NNM have similar performances and proposed DIP method is slightly better in the sense of target peaks and artifacts. 2D-SL0 has the worst performance among all and some artifacts can be observed around the target.

In the last case, missing ratio is increased to 70\% and this is the extreme case in our experiments. Generally, the performance of all methods deteriorated considerably as expected and RD imaging results clearly show the effect of the high missing ratio. The artifact levels in IALM result are significantly increased and target peaks are attenuated, however, the target is still distinguishable from the background. 2D-SL0 has many artifacts and these artifacts can be mixed with target peaks. Target structure is also corrupted. Among all the imaging methods, NNM has the worst performance for this missing ratio. The target structure is entirely lost and it has line artifacts starting from the center point. The proposed DIP method shows the best performance as it can be clearly observed from the visual results. There is some attenuation in target peaks however the structure is well preserved.

 When each method is examined individually, it is seen that IALM demonstrates a good performance up to a missing ratio of 70\%. Despite a significant increase in sidelobes and artifacts at this ratio, the target remains distinguishable from the background. Similarly, in 2D-SL0, the level of distortion increases with the missing ratio, and at a 70\% missing ratio the target structure was also affected and distorted. NNM performs well enough in the first two cases, however it produced the poorest result at a missing ratio of 70\%. The optimization algorithm for NNM operates patch-wise and always requires existing values in both vertical and horizontal directions. As a result, the patch size is increased at the 70\% missing ratio, and larger patches lead to a reduced performance for NNM, which results in its worst performance in the bottom image.The proposed DIP is the only method that remains effective at a 70\% missing ratio. It presents no deterioration in the first two cases, however has significant corruption in the extreme case of the pixel-wise missing data. The proposed DIP method outperforms the comparison methods and obtains better results visually. There is some attenuation in the target peaks due to the increasing missing ratio as expected. 
 
 Table \ref{tab:1} presents quantitative results and proves the superior reconstruction performance of the proposed DIP method compared to the other methods. The RMSE results increase from top to bottom as the level of corruption in the imaging results increases, while the correlation scores decrease as expected. The insufficient reconstruction performance leads to a smaller correlation value between the original and reconstructed images. In most cases, IC scores are coherent with the RMSE and Correlation scores, although in some cases, the IC scores are higher despite the poor visual results. As given before, comparing IC scores is only meaningful when RMSE and Correlation scores are close to each other. Otherwise, it may not be compatible with the visual results.

In column-wise (azimuth) missing case, the masks are presented in Fig. \ref{fig:9}(a) and  distortion occurs on the cross-range axis which can be observed in Fig. \ref{fig:9}(b), as  seen in RD imaging results. The missing columns in the azimuth axis of the data leads to the mixing and distortion of the target region on the cross-range axis. In the visual results, there aren't many artifacts that affect the entire background. However, compared to pixel-wise cases, side lobe effects on the target region are more severe. In column-wise case, the RD imaging results demonstrate significant distortion and a dramatic loss of target structure with the increasing missing ratio. The reconstruction results of the methods illustrated in Fig. \ref{fig:9}(c)--(f) will be examined in detail for each missing ratio.  

In 30\% missing ratio, RD imaging result has low level of sidelobes and it has corruption on both directions. The imaging result of IALM is similar to RD imaging result since it can not work on cases where a whole row or column is missing. The 2D-SL0 reduces corruption, but it also attenuates the target peaks, which is not desired. Although the visual results of both NNM and the proposed DIP are satisfactory, the imaging results of the proposed DIP provides better target peaks.

In the second case, 50\% missing ratio is applied to data in column-wise (azimuth) and a high level of corruption is observed in RD imaging result. IALM has the same visual result with RD imaging, as expected. Reconstruction performances are significantly degraded for 2D-SL0 and NNM, and target peaks are weaker for both methods. Proposed DIP has strong target peaks while also showing some unwanted artifacts on target region.

In the last case, 70\% missing ratio is applied and this is the extreme case for column-wise scenario. Missing data on the view angle results in significant corruption of the RD imaging result. The corruption on cross-range covers the target structure thus the target is not clearly detectable due to its distorted appearance. The IALM can not improve the quality of the corrupted image, as is it is clear from the visual results. Imaging results of 2D-SL0 and NNM are useless since target peaks are almost completely lost and structure of the target is highly corrupted. Between these two methods, NNM has the worst result. Among all methods, only the proposed DIP is capable of reconstructing a satisfactory target image with low level of artifacts. However, it does slightly attenuate the target peaks when compared to its previous two results.

\begin{figure*}[t!]
\begin{center}
\includegraphics[width=.68\textwidth]{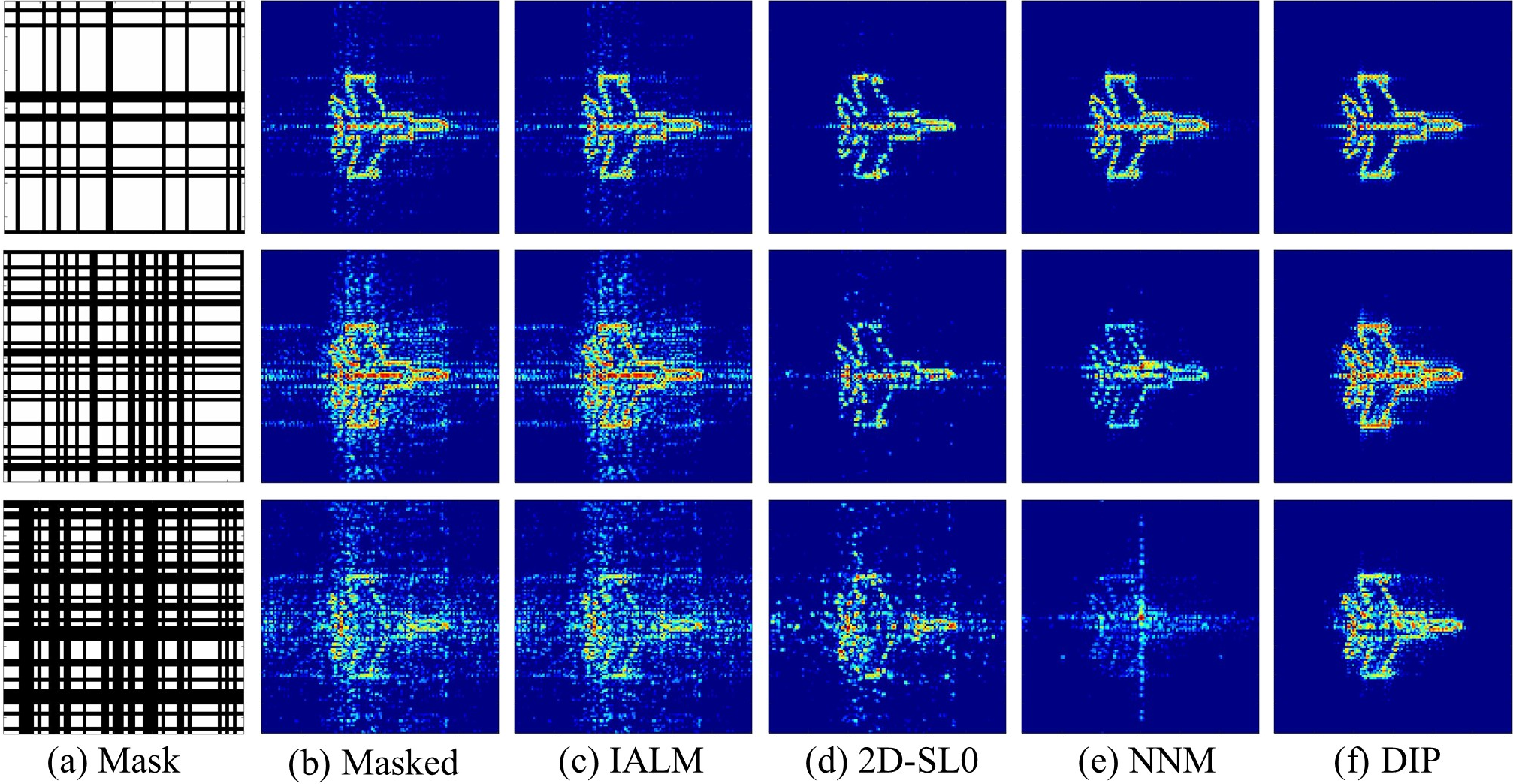}
\caption{Visual comparison of ISAR imaging completion methods for simulated Mig-25. The compressed rates (\%) are: 30, 50, and 70 from top to bottom. (a) Applied masks, (b) the corrupted ISAR image and the reconstruction results for (c) IALM, (d) 2D-SL0, (e) NNM, and (f) proposed DIP. Images are normalized to their maximum and top 20 dB are shown.}
\label{fig:10}       
\end{center}
\end{figure*}

\begin{table*}[!h]
\caption{Quantitative comparison results of ISAR imaging methods for simulated Mig-25 data in compressed case.}
\centering 
  \resizebox{13.5 cm}{!}{%
\begin{tabular}{c c c c c c c c c c c c} 
\toprule
\multirow{ 2}{*}{\textbf{Method}} &\multicolumn{3}{c}{\textbf{RMSE}} & & \multicolumn{3}{c}{\textbf{Correlation}} & &  \multicolumn{3}{c}{\textbf{Contrast}}\\
\cmidrule{2-4} 
\cmidrule{6-8}
\cmidrule{10-12}

  & L = 30\% & L = 50\% & L = 70\% & & L = 30\% & L = 50\% & L = 70\% & & L = 30\% & L = 50\% & L = 70\% \\ \cmidrule(r){1-12}

IALM	&	0.4405	&	0.7887	&	0.8540	&&	0.9073	&	0.8286	&	0.6604	&&	1.3458	&	1.2628	&	1.0157	\\
2D-SL0	&	0.4843	&	0.5736	&	0.7515	&&	0.8445	&	0.7717	&	0.6618	&&	1.5317	&	1.416	&	1.2329	\\
NNM	&	0.2439	&	0.5499	&	0.8145	&&	0.9648	&	0.8066	&	0.5528	&&	1.5654	&	1.4699	&	1.3671	\\
Proposed DIP	&	\textbf{0.1199}	&	\textbf{0.4002}	&	\textbf{0.4371}	&&	\textbf{0.9919}	&	\textbf{0.9440}	&	\textbf{0.8791}	&&	\textbf{1.6463}	&	\textbf{1.6451}	&	\textbf{1.5794}	\\

\hline
\end{tabular}
}
\label{tab:3} 
\end{table*}

If the reconstruction performances are compared for all missing ratios, IALM has the worst performance compared to the other methods since its reconstruction performance is significantly impacted by missing data in an entire column or row. The performance of 2D-SL0 is acceptable up to a missing ratio of 70\% missing ratio. However, the target is not  distinguishable from the background in the last case for the missing data. NNM produces satisfactory results for the first two cases, but it completely fails for the scenario of extreme missing ratio. Our proposed DIP method consistently achieves  competitive results among the all methods.

Quantitative results of column-wise case for simulated Mig-25 ISAR data can be seen on Table \ref{tab:2}. In terms of RMSE score, IALM almost has the worst results in all different missing ratios which can be seen clearly from the Table \ref{tab:2}. NNM outperforms 2D-SL0 in terms of the RMSE score for the first 2 cases, however in the last missing ratio, NNM attenuates the target peaks, resulting in an increase in its RMSE score. Correlation and Contrast results are decreased for all imaging methods with respect to increasing total number of missing samples as it is expected. These two metrics align well with the RMSE results in terms of quantitative performance for the column-wise missing scenario.

As a last scenario for missing data case, data compression (compressed case) is applied on simulated Mig-25 ISAR data. In the compressed case, random columns or rows are removed for the given missing case ratio. Different masks can be used to achieve the same compression rate, but the missing data ratio of columns and rows is adjusted to maintain the same ratio in the frequency and azimuth axes. The mask and the result of RD imaging for the compressed case is presented in Fig. \ref{fig:10}(a) and (b). The imaging results of the methods can be seen in Fig.~\ref{fig:10}(c)--(f). IALM shows poor performance when there is missing data in both an entire row and column. As mentioned before, IALM is unable to reconstruct data when whole row or column is missing. 2D-SL0 performs better in the first two cases, however it fails to preserve the target structure when the compression rate is 70\%. NNM obtains satisfactory results at a compression rate of 30\% which is the simplest case. In the 50\% and 70\% missing cases, while the NNM method successfully suppressed artifacts, however it also attenuates the target peaks. 

As in the other scenarios that was previously explained, the quantitative results in Table~\ref{tab:3} are compatible with the performance of the methods in the visual results. Table~\ref{tab:3}  shows that our proposed method has obtained the best results in terms of three metrics. NNM and 2D-SL0 have competitive results with DIP for the first two missing ratios however they failed in the extreme case. The worst method among them is IALM and this mainly reasoned by its optimization nature.

\begin{figure*}[h!]
\begin{center}
\includegraphics[width=0.75\textwidth]{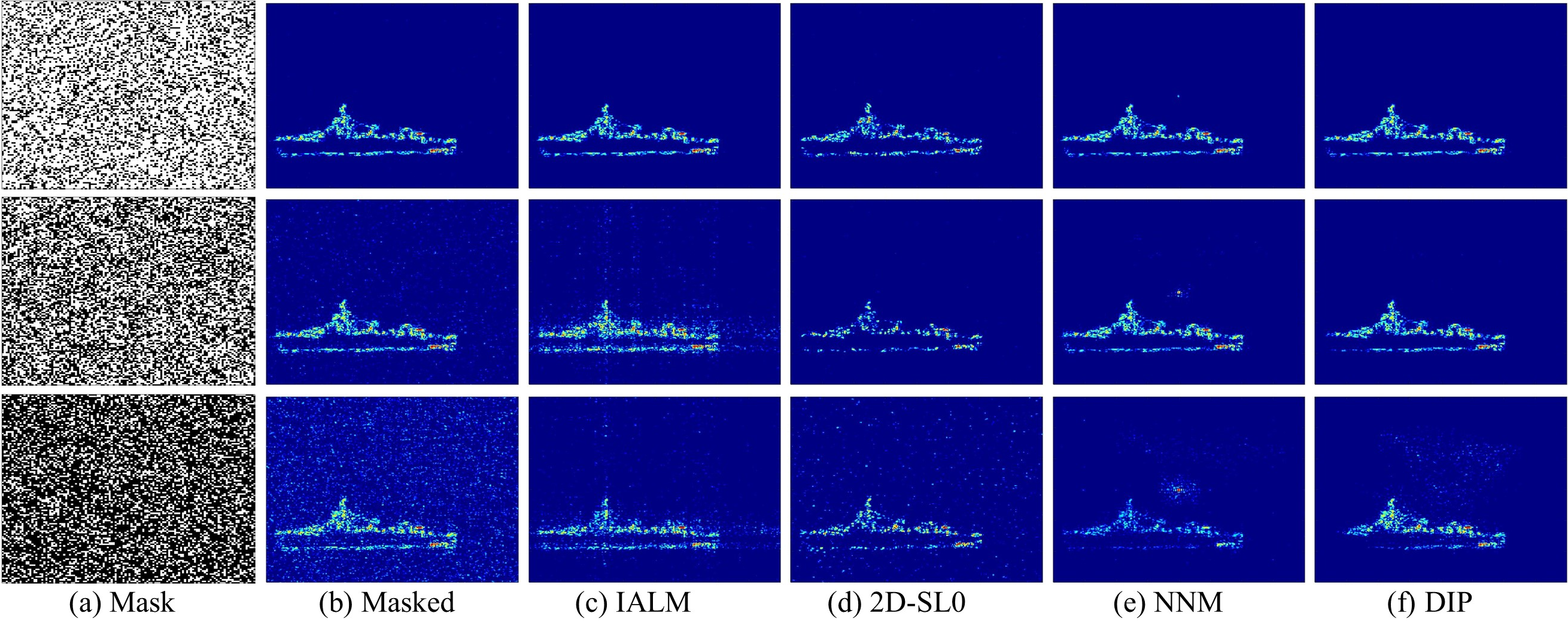}
\caption{Visual comparison of ISAR imaging completion methods for simulated Uss Fletcher warship. The pixel wise missing rates (\%) are: 30, 50, and 70 from top to bottom. (a) Applied masks, (b) the corrupted ISAR image and the reconstruction results for (c) IALM, (d) 2D-SL0, (e) NNM, and (f) proposed DIP. Images are normalized to their maximum and top 20 dB are shown.}
\label{fig:11}    
\end{center}
\end{figure*}

\begin{table*}[!h]
\caption{Quantitative comparison results of ISAR imaging methods for simulated USS FLETCHER warship data in pixel-wise missing case.} 
\centering
 \resizebox{13.5 cm}{!}{%
\begin{tabular}{c c c c c c c c c c c c} 
\toprule
\multirow{ 2}{*}{\textbf{Method}} &\multicolumn{3}{c}{\textbf{RMSE}} & & \multicolumn{3}{c}{\textbf{Correlation}} & &  \multicolumn{3}{c}{\textbf{Contrast}}\\
\cmidrule{2-4} 
\cmidrule{6-8}
\cmidrule{10-12}

  & L = 30\% & L = 50\% & L = 70\% & & L = 30\% & L = 50\% & L = 70\% & & L = 30\% & L = 50\% & L = 70\% \\ \cmidrule(r){1-12}

IALM	&	\textbf{0.2225}	&	0.6281	&	0.7804	&&	\textbf{0.9657}	&	0.7695	&	0.7270	&&	1.5011	&	1.1362	&	1.0398	\\
2D-SL0	&	0.5956	&	0.6285	&	0.8481	&&	0.7273	&	0.6858	&	0.5823	&&	1.4036	&	1.2421	&	1.0768	\\
NNM	&	0.2502	&	0.4540	&	0.6534	&&	0.9582	&	0.8881	&	0.6708	&&	1.4498	&	1.2574	&	1.0368	\\
Proposed DIP	&	0.2261	&	\textbf{0.3564}	&	\textbf{0.6385}	&&	0.9642	&	\textbf{0.9124}	&	\textbf{0.7372}	&&	\textbf{1.5227}	&	\textbf{1.4294}	&	\textbf{1.3157}	\\

\hline
\end{tabular}
}
\label{tab:4} 
\end{table*}

\begin{figure*}[h!]
\begin{center}
\includegraphics[width=0.85\textwidth]{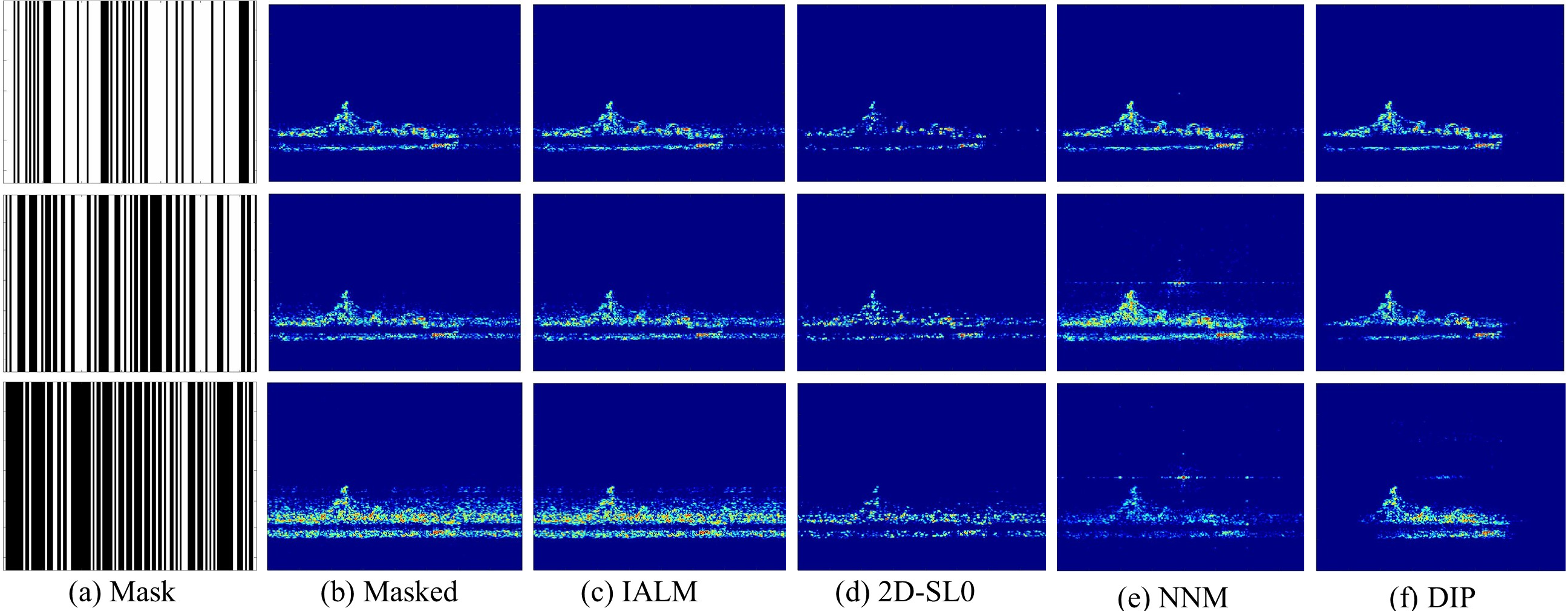}
\caption{Visual comparison of ISAR imaging completion methods for simulated Uss Fletcher warship. The column-wise missing rates (\%) are: 30, 50, and 70 from top to bottom. (a) Applied masks, (b) the corrupted ISAR image and the reconstruction results for (c) IALM, (d) 2D-SL0, (e) NNM, and (f) proposed DIP. Images are normalized to their maximum and top 20 dB are shown.}
\label{fig:12}       
\end{center}
\end{figure*}

\begin{table*}[!h]
\caption{Quantitative comparison results of ISAR imaging methods for simulated USS FLETCHER warship data in column-wise missing case.}
\centering 
 \resizebox{13.5 cm}{!}{%
\begin{tabular}{c c c c c c c c c c c c} 
\toprule
\multirow{ 2}{*}{\textbf{Method}} &\multicolumn{3}{c}{\textbf{RMSE}} & & \multicolumn{3}{c}{\textbf{Correlation}} & &  \multicolumn{3}{c}{\textbf{Contrast}}\\
\cmidrule{2-4} 
\cmidrule{6-8}
\cmidrule{10-12}

  & L = 30\% & L = 50\% & L = 70\% & & L = 30\% & L = 50\% & L = 70\% & & L = 30\% & L = 50\% & L = 70\% \\ \cmidrule(r){1-12}
IALM	&	0.4601	&	0.6757	&	1.2345	&&	0.8812	&	0.7777	&	0.6495	&&	1.4582	&	1.4364	&	1.4316	\\
2D-SL0	&	0.5973	&	0.7035	&	0.8169	&&	0.7337	&	0.6605	&	0.5755	&&	\textbf{1.5775}	&	\textbf{1.5394}	&	\textbf{1.5272}	\\
NNM	&	0.3714	&	0.5379	&	\textbf{0.6749}	&&	0.9196	&	0.8270 &	0.6251	&&	1.4775	&	1.4027	&	1.2618	\\
Proposed DIP	&	\textbf{0.3084}	&	\textbf{0.4508}	&	0.7341	&&	\textbf{0.9372}	&	\textbf{0.8550}	&	\textbf{0.6738}	&&	1.5152	&	1.4845	&	1.4636	\\

\hline
\end{tabular}
}
\label{tab:5} 
\end{table*}

\begin{figure*}[h!]
\begin{center}
\includegraphics[width=0.85\textwidth]{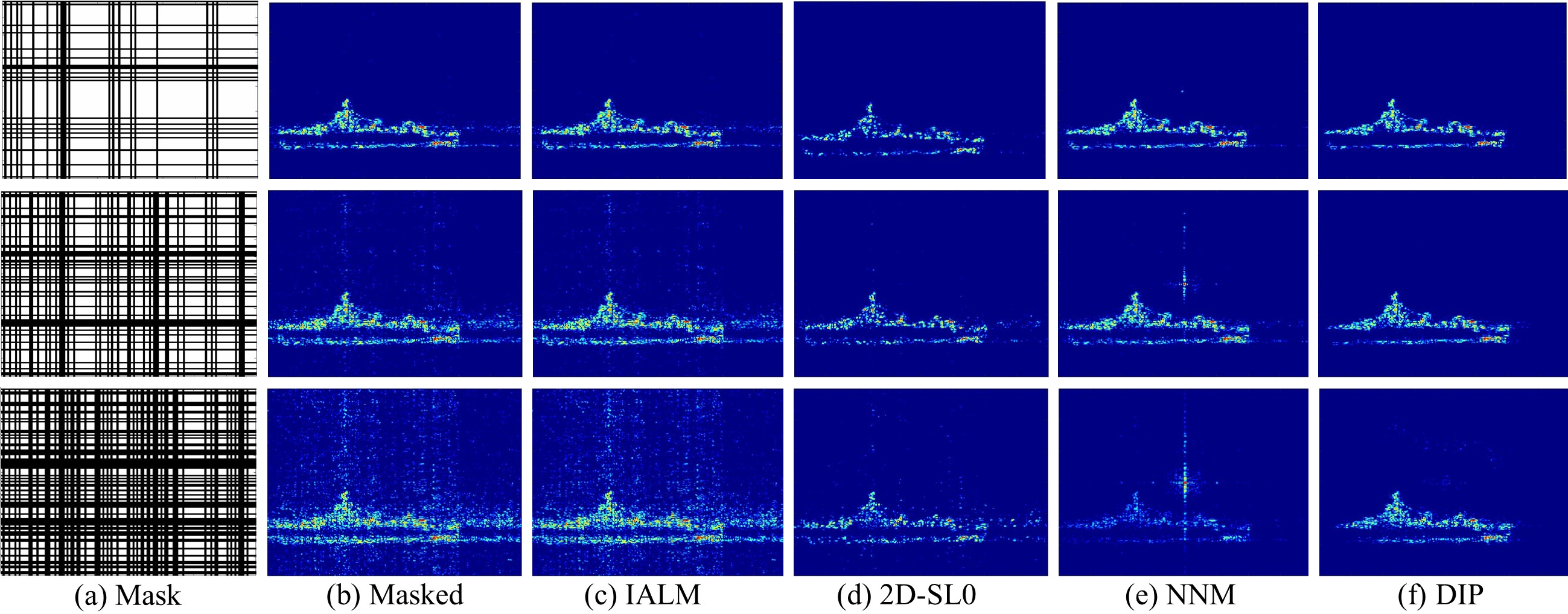}
\caption{Visual comparison of ISAR imaging completion methods for simulated Uss Fletcher warship. The compressed rates (\%) are: 30, 50, and 70 from top to bottom. (a) Applied masks, (b) the corrupted ISAR image and the reconstruction results for (c) IALM, (d) 2D-SL0, (e) NNM, and (f) proposed DIP. Images are normalized to their maximum and top 20 dB are shown.}
\label{fig:13}      
\end{center}
\end{figure*}

\begin{table*}[!h]
\caption{Quantitative comparison results of ISAR imaging methods for simulated USS FLETCHER warship data in compressed case.} 
\centering 
 \resizebox{13.5 cm}{!}{%
\begin{tabular}{c c c c c c c c c c c c} 
\toprule
\multirow{ 2}{*}{\textbf{Method}} &\multicolumn{3}{c}{\textbf{RMSE}} & & \multicolumn{3}{c}{\textbf{Correlation}} & &  \multicolumn{3}{c}{\textbf{Contrast}}\\
\cmidrule{2-4} 
\cmidrule{6-8}
\cmidrule{10-12}

  & L = 30\% & L = 50\% & L = 70\% & & L = 30\% & L = 50\% & L = 70\% & & L = 30\% & L = 50\% & L = 70\% \\ \cmidrule(r){1-12}
IALM	&	0.5215	&	0.8498	&	1.2634	&&	0.9191	&	0.8464	&	0.7650	&&	1.3126	&	1.1428	&	0.9877	\\
2D-SL0	&	0.5869	&	0.6544	&	0.7412	&&	0.8151	&	0.7870	&	0.7419	&&	1.5041	&	1.3737	&	1.2352	\\
NNM	&	0.3292	&	0.5524	&	0.7590	&&	0.9542	&	0.8876	&	0.7187	&&	1.4638	&	1.3138	&	1.2156	\\
Proposed DIP	&	\textbf{0.2653}	&	\textbf{0.4020}	&	\textbf{0.5397}	&&	\textbf{0.9645}	&	\textbf{0.9232}	&	\textbf{0.8589}	&&	\textbf{1.5107}	&	\textbf{1.4635}	&	\textbf{1.3861}	\\

\hline
\end{tabular}
}
\label{tab:6} 
\end{table*}

Simulated USS Fletcher data is also used for further analysis of the methods. Same missing scenarios with same missing ratios are applied. Visual results are presented in Figs. \ref{fig:11}, \ref{fig:12} and \ref{fig:13} for pixel-wise, column-wise and compressed cases, respectively. Same as before 30\%, 50\% and 70\% missing ratios are applied and quantitative results can be seen in Tables \ref{tab:4}, \ref{tab:5} and \ref{tab:6}, respectively. Performances of the methods are similar with the Mig-25 data in most of the cases. Despite slight differences, it can be observed that proposed DIP achieves the best reconstruction performance both visually and quantitatively. The worst performance is obtained by IALM since it fails the last two scenarios.

\begin{figure*}[!h]

\begin{center}
\includegraphics[width=0.7\textwidth]{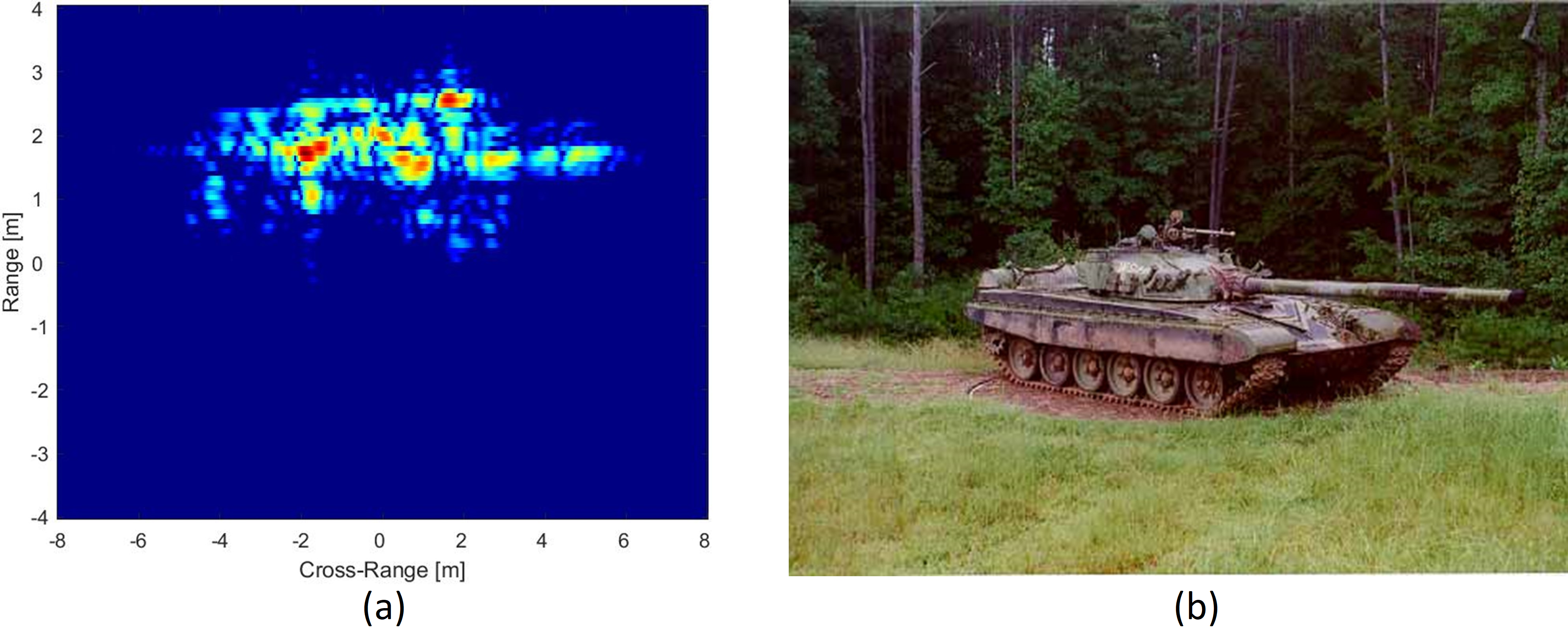}
\caption{(a) ISAR imaging result of real T-72 Tank with full data by using RD algorithm. ISAR image is normalized to the maximum amplitude and
shown on a top 30 dB; (b) Photo of T-72 Tank.}
\label{fig:14}       
\end{center}
\end{figure*}

\begin{table}[!h]
\centering
\caption{Details of real data.}
\label{tab:real_data}
\resizebox{0.4\textwidth}{!}{%
\begin{tabular}{|l|c|}
\hline
File Name                & f111HH21 \\ \hline
Carrier Frequency ($f_0$)   & 9.6 GHz  \\ \hline
Frequency Step           & 3 MHz    \\ \hline
Rotation Angle Step      & 0.05     \\ \hline
Aspect Angle             & 89.2361  \\ \hline
Total Frequency Step (N) & 221      \\ \hline
Total Aperture Step (M)  & 79       \\ \hline
\end{tabular}%
}
\end{table}

\subsection{Real Dataset Results}

In order to prove robustness and general application of  the proposed model, real data has also been used for detail analysis. As  real data, turntable ISAR data of T-72 tank target are selected. The data  are collected by Georgia Tech Research Institute (GTRI).  It is one of the open source ISAR real data used in many studies \cite{Sonia_2016,Qui,Sevket_hoca}. The data set is provided by Air Force Research Laboratory
(AFRL) website and can be freely reached \cite{AFRL}. The ISAR image of the real data and its real
image are presented in Fig.~\ref{fig:14}.

The GTRI turntable dataset contains multi-polarized X-band ISAR data which has 9.6 GHz center frequency. There are four types of polarization such as HH, HV, VH and VV for transmitter/receiver couples. In total, 6 folders contain 29 sub folders and each sub folder contains 340 files that have 4 different polarizations of 85 files. Each file represents azimuth steps thus full-aperture 3.9° is divided into the 85 equal step with 0.05° per step. Frequency step is 3 MHz for center frequency 9.6 GHz and frequency changes between 9.27 GHz to 9.93 GHz. The original data has higher range and cross-range resolution but provided dataset is downsampled to the 1 foot x 1 foot range and cross-range resolution. The data also provides depression angles so the given data set has the capability to generate 3-D ISAR images. Details are given in Table \ref{tab:7}.

\begin{figure*}[!h]
\begin{center}
\includegraphics[width=0.9\textwidth]{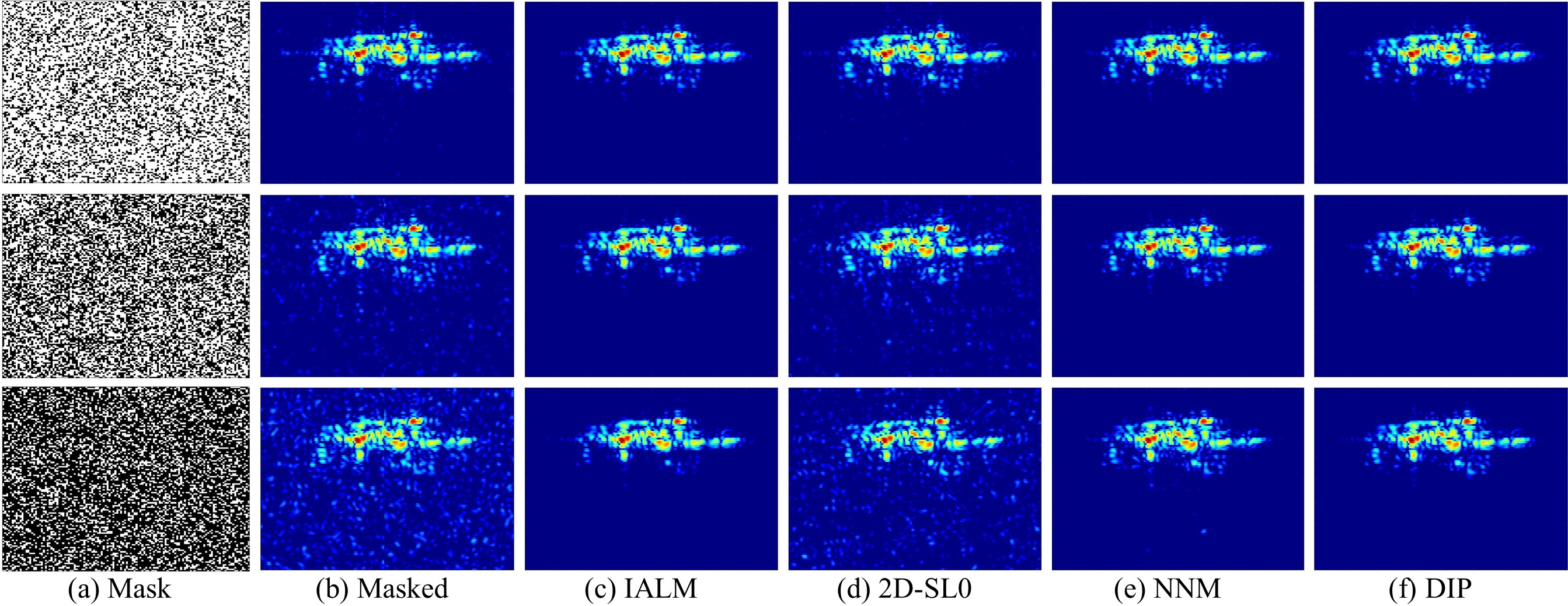}
\caption{Visual comparison of ISAR imaging completion methods for real T-72 tank. The pixel-wise missing rates (\%) are: 30, 50, and 70 from top to bottom. (a) Applied masks, (b) the corrupted ISAR image and the reconstruction results for (c) IALM, (d) 2D-SL0, (e) NNM, and (f) proposed DIP. Images are normalized to their maximum and top 30 dB are shown.}
\label{fig:15}      
\end{center}
\end{figure*}

\begin{table*}[!h]
\caption{Quantitative comparison results of ISAR imaging methods for real data in pixel-wise missing case.} 
\centering 
 \resizebox{13.5 cm}{!}{%
\begin{tabular}{c c c c c c c c c c c c} 
\toprule
\multirow{ 2}{*}{\textbf{Method}} &\multicolumn{3}{c}{\textbf{RMSE}} & & \multicolumn{3}{c}{\textbf{Correlation}} & &  \multicolumn{3}{c}{\textbf{Contrast}}\\
\cmidrule{2-4} 
\cmidrule{6-8}
\cmidrule{10-12}

  & L = 30\% & L = 50\% & L = 70\% & & L = 30\% & L = 50\% & L = 70\% & & L = 30\% & L = 50\% & L = 70\% \\ \cmidrule(r){1-12}

IALM	&	0.1308	&	0.1401	&	0.1782	&&	0.9931	&	0.9917	&	0.9852	&&	1.9074	&	1.8754	&	1.8285	\\
2D-SL0	&	0.3146	&	0.4324	&	0.9256	&&	0.9531	&	0.9261	&	0.7398	&&	1.7814	&	1.6797	&	1.2341	\\
NNM	&	0.1333	&	0.1459	&	0.2204	&&	0.9925	&	0.9902	&	0.9805	&&	1.9052	&	1.9006	&	1.7838	\\
Proposed DIP	&	\textbf{0.1304}	&	\textbf{0.1348}	&	\textbf{0.1404}	&&	\textbf{0.9937}	&	\textbf{0.9920}	&	\textbf{0.9916}	&&	\textbf{1.9083}	&	\textbf{1.8761}	&	\textbf{1.8615}	\\

\hline
\end{tabular}
}
\label{tab:8} 
\end{table*}

\begin{figure*}[h!]
\begin{center}
\includegraphics[width=0.9\textwidth]{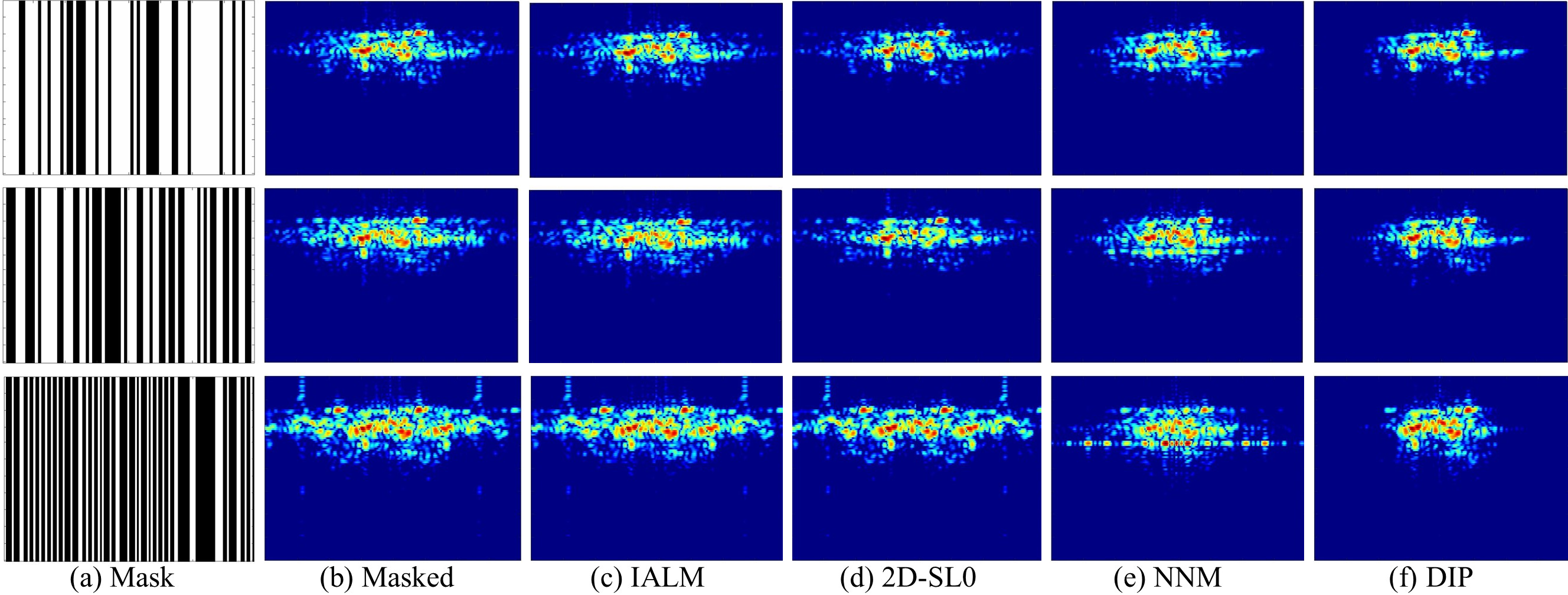}
\caption{Visual comparison of ISAR imaging completion methods for real T-72 tank. The column-wise missing rates (\%) are: 30, 50, and 70 from top to bottom. (a) Applied masks, (b) the corrupted ISAR image and the reconstruction results for (c) IALM, (d) 2D-SL0, (e) NNM, and (f) proposed DIP. Images are normalized to their maximum and top 30 dB are shown.}
\label{fig:16}       
\end{center}
\end{figure*}

\begin{table*}[!h]
\caption{Quantitative comparison results of ISAR imaging methods for real data in column-wise missing case.} 
\centering 
 \resizebox{13.5 cm}{!}{%
\begin{tabular}{c c c c c c c c c c c c}
\toprule
\multirow{ 2}{*}{\textbf{Method}} &\multicolumn{3}{c}{\textbf{RMSE}} & & \multicolumn{3}{c}{\textbf{Correlation}} & &  \multicolumn{3}{c}{\textbf{Contrast}}\\
\cmidrule{2-4} 
\cmidrule{6-8}
\cmidrule{10-12}

  & L = 30\% & L = 50\% & L = 70\% & & L = 30\% & L = 50\% & L = 70\% & & L = 30\% & L = 50\% & L = 70\% \\ \cmidrule(r){1-12}

IALM	&	0.35991	&	0.52907	&	0.8681	&&	0.95302	&	0.91064	&	0.85175	&&	1.8625	&	1.8415	&	1.8379	\\
2D-SL0	&	0.38488	&	0.53608	&	0.8836	&&	0.93857	&	0.89083	&	0.83818	&&	1.8583	&	1.843	&	1.8397	\\
NNM	&	0.29447	&	0.51581	&	1.025	&&	0.96399	&	0.91872	&	0.72553	&&	1.8936	&	1.8813	&	1.8782	\\
Proposed DIP	&	\textbf{0.16282}	&	\textbf{0.27519}	&	\textbf{0.51387}	&&	\textbf{0.98808}	&	\textbf{0.97181}	&	\textbf{0.91872}	&&	\textbf{1.9022}	&	\textbf{1.8985}	&	\textbf{1.8929}	\\

\hline
\end{tabular}
}
\label{tab:9} 
\end{table*}

The ISAR data of real T-72 tank is tested under three different missing scenarios. As a first scenario, pixel-wise missing case is applied on raw ISAR data. As it can be seen from Fig. \ref{fig:15}(a) and (b), different missing ratios are used such as 30\%, 50\% and 70\%, respectively. From left to right, imaging results of the methods are presented in Fig. \ref{fig:15}(c)--(f). For the first pixel-wise missing case, 30\% missing ratios is used and this is the simplest case. The RD imaging result presents a few number of artifacts  randomly distributed around the target. All the methods provide visually satisfactory results with very few artifacts on random locations and it is hard to detect them by visual inspection. For the 50\% missing ratio, artifacts are significantly increased on RD imaging result. All imaging methods except the 2D-SL0 remove significant amount of the artifacts and visual results are almost same for all.  For 2D-SL0, the artifacts can be seen by visual inspection  below  the reconstructed image. In the final case, the extreme missing ratio 70\% is applied and it highly corrupts the RD imaging result. It can be clearly observed that artifacts are randomly distributed all over the image and some artifacts are as strong as target peaks thus some of the peaks that  belong to target structure are mixed with undesired artifacts. 2D-SL0 reduces artifacts however it fails to remove all of them. IALM, NNM and proposed DIP methods visually separated from 2D-SL0 in positive way at missing ratio 70\%. IALM removes all artifacts while it also attenuates and removes some target peaks as it can be observed visually. NNM reconstructs the target structure however some artifacts  which are located far from the target vicinity still remain. The  proposed DIP method consistently achieves  competitive results for all levels of missing ratio. Quantitative results are given in Table \ref{tab:8} and show that RMSE and Correlation scores are similar for IALM, NNM and proposed DIP. In the extreme missing ratio of 70\%, 2D-SL0 can lead to an undesirable attenuation in target peaks. This causes a significant decrease in the Correlation score. If simulated and real data results are compared in detail, the results obtained from real data have significant positive impact in the reconstruction performance of the methods in pixel-wise case. The benefits of such an impact can be seen from both visual and quantitative results. Among all the visual results of the imaging methods, proposed DIP has superior performance since it removes all artifacts visually while preserving target peaks, therefore it has best quantitative results in terms of RMSE and Correlation scores.

 As observed earlier, column-wise missing was a challenging case. Real ISAR data are also tested with column-wise (azimuth) missing case to prove superiority of the proposed DIP method. Same as before, three different missing ratios are applied such as 30\%, 50\% and 70\%, the masks and corresponding RD imaging results can be seen in Fig. \ref{fig:16}(a) and (b), respectively. The reconstruction results of the methods are presented in Fig. \ref{fig:16}(c)--(f). For 30\% missing ratio, Fig.~\ref{fig:16} illustrates that the target starts to spread along the cross-range axis. As previous experiments have shown, column-wise missing can cause significant distortion along the cross-range axis. IALM performs almost the same visual reconstruction quality with RD imaging result. 2D-SL0 reduces sidelobes that are located along the cross-range axis, while preserving the target structure. Compared to 2D-SL0, NNM reduces sidelobes more effectively, however it mixes the target structure in some regions. Among all, the proposed DIP achieves the best visual results without any artifacts while some of the target peaks become slightly weaker. For 50\% missing ratio RD imaging result has huge corruption along the cross-range axis and target structure is hardly recognizable by visual inspection. Imaging result of IALM is almost same with  RD imaging and it does not make any contribution to the visual results. 2D-SL0 reduces some of the sidelobes, however target structure is also mixed. NNM successfully removes sidelobes but it does not reconstruct target structure properly. The proposed DIP removes sidelobes for this missing ratio and preserves  the target structure better compared to other methods. Since missing columns are selected randomly during the creation of the masks, some of these columns may be concatenated randomly and form block-wise missing regions. This phenomenon can have a significant impact on the reconstruction performance. In column-wise scenario, 70\% missing ratio is also applied to real data as an extreme case. As it can be seen from figure RD imaging result has lost target signature completely. Target structure and target location are not detectable  by visual inspection due to the high level of sidelobes. Imaging results of IALM and 2D-SL0 are also mixed on cross-range axis, same as RD imaging result. In both methods, target structure and target location are not detectable visually. NNM has noticeable artifacts on cross-range axis that are not associated with the target structure. Artifacts can be observed around the middle of the reconstructed image and reconstructed target peaks do not overlap with the actual target peaks, resulting in an undetectable target signature. The proposed DIP effectively eliminates all sidelobes, however there are some missing parts of the targets, which leads to  unsatisfactory reconstruction results. RMSE and Correlation values are given in Table \ref{tab:9} and it can be seen that reconstruction performances reduce significantly with respect to increasing missing ratios. Our proposed method gives the best performance among all in terms of quantitative metrics. NNM method follows the DIP method except the extreme missing ratio. IALM and 2D-SL0 have almost no difference at a noticeable level.

\begin{figure*}[h!]
\begin{center}
\includegraphics[width=0.9\textwidth]{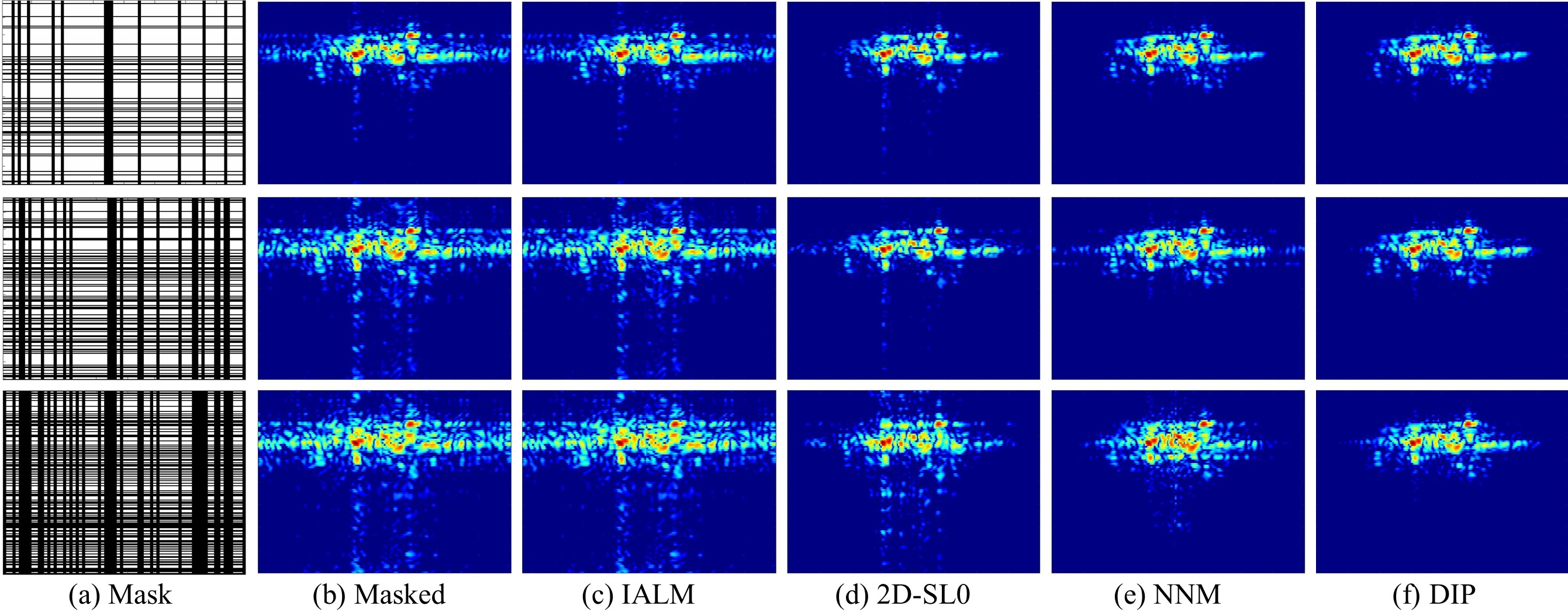}
\caption{Visual comparison of ISAR imaging completion methods for real T-72 tank. The compressed rates (\%) are: 30, 50, and 70 from top to bottom. (a) Applied masks, (b) the corrupted ISAR image and the reconstruction results for (c) IALM, (d) 2D-SL0, (e) NNM, and (f) proposed DIP. Images are normalized to their maximum and top 30 dB are shown.}
\label{fig:17}       
\end{center}
\end{figure*}

\begin{table*}[!h]
\caption{Quantitative comparison results of ISAR imaging methods for real data in compressed case.} 
\centering 
\resizebox{13.5 cm}{!}{%
\begin{tabular}{c c c c c c c c c c c c} 
\toprule
\multirow{ 2}{*}{\textbf{Method}} &\multicolumn{3}{c}{\textbf{RMSE}} & & \multicolumn{3}{c}{\textbf{Correlation}} & &  \multicolumn{3}{c}{\textbf{Contrast}}\\
\cmidrule{2-4} 
\cmidrule{6-8}
\cmidrule{10-12}

  & L = 30\% & L = 50\% & L = 70\% & & L = 30\% & L = 50\% & L = 70\% & & L = 30\% & L = 50\% & L = 70\% \\ \cmidrule(r){1-12}

IALM	&	0.4974	&	0.9547	&	1.3494	&&	0.9181	&	0.7924	&	0.6947	&&	1.6398	&	1.4462	&	1.2817	\\
2D-SL0	&	0.4570	&	0.5974	&	0.9129	&&	0.9264	&	0.8540	&	0.7274	&&	1.8348	&	1.7264	&	1.5905	\\
NNM	&	0.1433	&	0.2820	&	0.5632	&&	0.9893	&	0.9648	&	0.9024	&&	1.8606	&	1.8316	&	1.7543	\\
Proposed DIP	&	\textbf{0.0791}	&	\textbf{0.1102}	&	\textbf{0.23393}	&&	\textbf{0.99679}	&	\textbf{0.99364}	&	\textbf{0.9713}	&&	\textbf{1.8831}	&	\textbf{1.8739}	&	\textbf{1.8669}	\\

\hline
\end{tabular}
}
\label{tab:10} 
\end{table*}

The final missing scenario includes the evaluation of the compressed case with three different  missing ratios identical to those used in previous experiments. Masks and corresponding RD imaging results are presented in Fig. \ref{fig:17}(a) and (b). Visual performance of the methods are presented in  Fig. \ref{fig:17}(c)--(f). The 30\% missing case is our first and simplest experiment for reconstruction performance evaluation. IALM has almost the same performance as the RD imaging result, which has some artifacts and sidelobes that are located in both range and cross-range axes. 2D-SL0 removes sidelobes that are located on cross-range direction but it has some artifacts that are located along the range axis and target peaks are attenuated in some parts of the reconstructed image. NNM successfully removes all artifacts and sidelobes in two axes, and target peaks are  approximately at the same level as the original image. The proposed DIP also removes all artifacts and sidelobes without attenuating target peaks and preserves the overall target structure effectively. As a second experiment for compressed case, 50\% missing ratio is utilized. Target structure and location is not clear in this case for compressed RD image. IALM method  performs poorly on this missing level and does not contribute to the results observed in the RD imaging. 2D-SL0 eliminates a high amount of the artifacts and sidelobes but there are still some artifacts that are located along the range and cross-range axes. Despite this, the target is still detectable. NNM removes sidelobes in cross-range axis but it has some artifacts that are located in the range axis. Target peaks remain same as the original and the target is clearly distinguishable from its surrounding background. The proposed DIP removes all the artifacts and sidelobes successfully and provides a result very similar to that obtained by RD  with full use of the data.  Among all, the proposed DIP method has clearly superior performance compared to others and NNM follows it. As the last missing ratio, 70\% compression is applied which is the extreme case and corruption of the RD imaging and IALM are increased. 2D-SL0 removes artifacts significantly, but target peaks are highly attenuated thus the target structure is not preserved as desired and cannot be clearly seen through visual inspection. Despite high number of missing samples along the cross-range axis, NNM is still effective in reducing sidelobes, while also retaining most of the target components. However, it still removes a few parts of the target. The proposed DIP removes all artifacts and sidelobes while it preserves the target peaks which can be observed visually. Among all methods, DIP takes the first place and NNM follows it. IALM has the worst results as expected. NNM and proposed DIP provided lower RMSE and higher Correlation when compared to the other imaging methods and it can be seen on Table \ref{tab:6} for all cases. IALM has the worst results among all in terms of quantitative metrics. The proposed DIP method achieves the best scores by far for all missing ratios in terms of different metrics as it can be seen in Table \ref{tab:10}.

\begin{figure}[h!]
\centering
\begin{center}

\includegraphics[width=.5\textwidth]{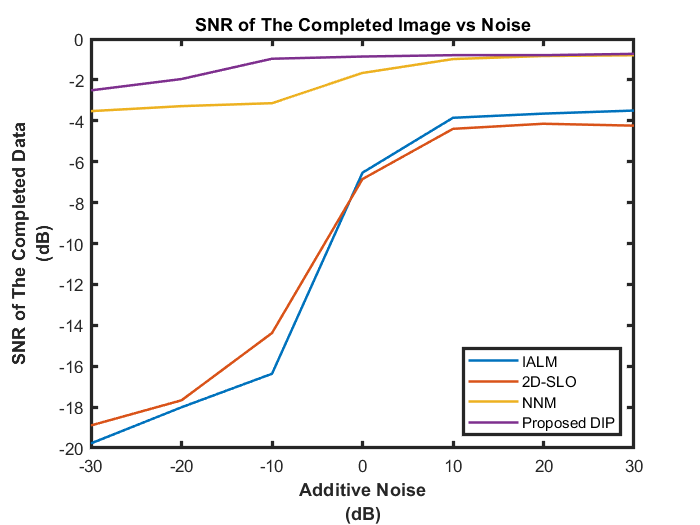
}

\caption{SNR of the imaging results in different noise level.}
\label{fig:18}       
\end{center}
\end{figure}

\subsection{Imaging Performance of The Methods in Noisy Case}

In real world applications noise is another challenge for imaging methods. Since  performance of the methods can  significantly decrease for low signal to noise ratio (SNR) levels, thus,  the imaging methods are evaluated in different noise levels. Simulated Mig-25 dataset is selected with pixel-wise missing at the missing ratio of 50\%. SNR values are set between -30dB to 30dB with steps of 10dB. It can be observed from fig 18 that the performance of compressive methods IALM and 2D-SLO deteriorate with the decreasing SNRs while low rank completion method NNM and our proposed method are not affected from noise.

The methods are tested in different noise level and SNR results are presented in Fig.~\ref{fig:18}. It
is seen that, the performance of the proposed method is more robust in terms of reconstruction performance in different noise level.

\subsection{Time Efficiency of Imaging Methods}
Besides  the signal reconstruction performance, time efficiency of the imaging methods is also crucial for real world implementations. To investigate the running time of the imaging methods, they are tested under same conditions. As it is seen in Table \ref{tab:11}, IALM was the faster one. 2D-SL0 was the second one, however it also requires dictionary based completion on sparse coefficients thus it may take longer time according to the matrix size. NNM is used with real data normally so complex ISAR raw data are separated into two parts as real and imaginary coefficient and it takes longer to run on ISAR image. Same as NNM, proposed DIP is applied on real and imaginary coefficients separately so it is the relatively slow when compared to the other imaging methods.  

\begin{table}[!h]
\caption{Running time comparison of the methods.} 
\centering
\resizebox{8.2 cm}{!}{%
\begin{tabular}{@{}ccccc@{}}
\toprule
Method & IALM & 2D-SL0 & NNM & Proposed DIP \\ \midrule
Running Time (s) & 1.94 & 2.54 & 4.16 & 7.097 \\ \bottomrule
\end{tabular}
}
\label{tab:11}
\end{table}

The iteration number is set to 1e4 for comparison methods, and tolerance values are utilized as 1e-5, thus the methods may converge earlier than the defined maximum number of iteration steps. In order to implement the same early stopping mechanism on the proposed DIP, which iteratively works on a DNN, related metrics are calculated for every iteration. Apart from the early stopping of DNN, metrics are controlled in the early stopping mechanism.
\begin{figure}[h!]
\centering
\begin{center}

\includegraphics[width=.48\textwidth]{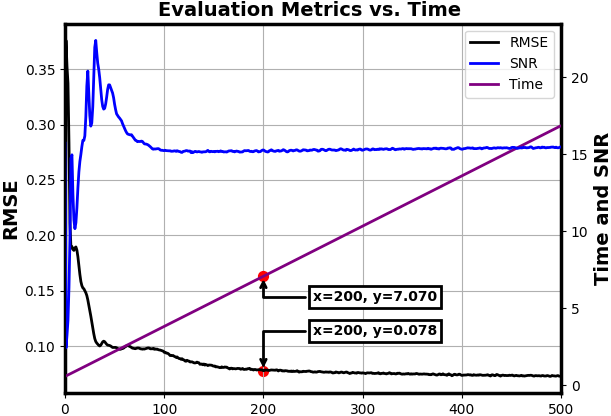}

\caption{An example speed test of proposed model for Time, RMSE and SNR versus iteration.}

\label{fig:19}       
\end{center}
\end{figure}

As it can be seen in Fig. \ref{fig:19} RMSE curve  converges approximately  at iteration 200, and the total elapsed time is approximately 7s at this point. It is also shown that the SNR curve  converges much before the RMSE curve at iteration 100. Thus, in this study, SNR value is checked to implement early stopping for the deep learning model. During the training process, if there are three consecutive instances of 1\% or less improvement in the signal-to-noise ratio (SNR), it is assumed that the SNR value has converged and early stop function terminates the training. Since iteration steps are equal, the total elapsed time for the 100th iteration is approximately 3.5s. Real and imaginary parts are reconstructed separately so that the total running time is equal to 7.097 as it is given in Fig. \ref{fig:19}. Imaging methods were tested for all loss cases on different ISAR data, and the results show the superiority of the proposed model, so a small difference in running time is not significantly important.

\section{Conclusion}

A complex data recovery method based on DIP has been introduced. The raw radar echo matrix is split into its real and imaginary parts which are  completed by separable DIP structures. The recovered components are then put together and the radar image is provided by the Fourier transform of the completed complex data. The method does not need any training. Gaussian noise is fed to the DIP structure, the missing data matrix is used as reference and the DIP structure tries to approximate the reference image in its iterations using fixed weight networks as priors. The proposed method does not suffer from the scatterer splitting problem encountered in compressive sensing based ISAR imaging methods and does not require any pre-processing step such as dictionary learning. It is robust to high missing ratios unlike the low rank based matrix completion approaches.  Results obtained for both simulated and real data demonstrate that the proposed method provides a good alternative for missing data case since it successfully completed complex missing data for even high missing ratios. Moreover, the proposed method will give good performance not only in ISAR but also in different missing data situations, especially working with complex data.


\end{document}